\documentclass[preprint]{aastex63}
\usepackage{natbib,amsmath,scalerel,placeins}

\usepackage{lmodern}
\usepackage[T1]{fontenc}
\usepackage{enumerate}

\shorttitle{Ionized disk in G345.49+1.47}
\newcommand{\hii}{H\textsc{ii}}
\newcommand{\hrl}{H38$\beta$}
\newcommand{\Msun}{$M_\sun$}
\newcommand{\hmyso}{G345.49+1.47}
\newcommand{\kms}{km~s$^{-1}$}
\newcommand{\los}{\mbox{l.o.s.}}
\newcommand{\gds}{G17.64+0.16}
\newcommand{\rB}{r_{\!\scriptscriptstyle B}}

\newcommand{\cs}{\reflectbox{\textsf{\small S}}}
\newcommand{\csreflect}{\textsf{\small S}}
\newcommand{\sigmatD}{\sigma_{\!\scaleto{3D}{4.5pt}}}
\newcommand{\chg}[1]{#1}
\newcommand{\cchg}[1]{#1}

\begin{document}

\title{A photoionized accretion disk around a young high-mass star.}
 
\author[0000-0003-0990-8990]{Andr\'es E. Guzm\'an}
\affiliation{National Astronomical Observatory of Japan, National Institutes of Natural Sciences, 2-21-1 Osawa, Mitaka, Tokyo 181-8588, Japan}

\author[0000-0002-7125-7685]{Patricio Sanhueza}
\affiliation{National Astronomical Observatory of Japan, National Institutes of Natural Sciences, 2-21-1 Osawa, Mitaka, Tokyo 181-8588, Japan}
\affiliation{Department of Astronomical Sciences, SOKENDAI (The Graduate University for Advanced Studies), 2-21-1 Osawa, Mitaka, Tokyo 181-8588, Japan}

\author[0000-0003-2343-7937]{Luis Zapata}
\affiliation{Instituto de Radioastronom\'ia y Astrof\'isica, Universidad Nacional Aut\'onoma de M\'exico, P.O. Box 3-72, 58090, Morelia, Michoac\'an, M\'exico}

\author{Guido Garay}
\affiliation{Departamento de Astronom\'ia, Universidad de Chile, Camino el Observatorio 1515, Las Condes, Santiago, Chile}

\author[0000-0003-2737-5681]{Luis Felipe Rodr\'iguez}
\affiliation{Instituto de Radioastronom\'ia y Astrof\'isica, Universidad Nacional Aut\'onoma de M\'exico, P.O. Box 3-72, 58090, Morelia, Michoac\'an, M\'exico}

\begin{abstract}
We present  high spatial resolution ($52$ au) observations of the high-mass young stellar object (HMYSO) G345.4938+01.4677 made using the Atacama Large Millimeter/sub-millimeter Array (ALMA).  This O-type HMYSO is located at 2.38 kpc and it is associated with a luminosity of $1.5\times10^5 L_\sun$. We detect circumstellar emission from the H$38\beta$ hydrogen recombination line showing a compact structure rotating  perpendicularly to the previously detected radio jet. We interpret this emission as tracing a photo-ionized accretion disk around the HMYSO. While this disk-like structure seems currently too small to sustain continued accretion, the data  present direct observational evidence of how disks can effectively survive the photo-ionization feedback from young high-mass stars. We also report the detection of a low-mass young stellar object in the vicinity of the HMYSO and suggest that it forms a high-mass and low-mass star binary system.
\end{abstract} 

\section{Introduction.}

In the recent years, and with the advent of new observational instruments such as the 
Atacama Large Millimeter/sub-millimeter Array (ALMA),
evidence have been gathering pointing toward  most high-mass stars acquiring their mass through 
disk accretion \citep{Tan2014prpl}. 
It is becoming apparent that  the principal characteristic which allows  accretion to proceed 
onto  most young high-mass young stellar objects (HMYSOs) --- despite the venerable  theoretical problems
of radiation and ionization feedback \citep{Wolfire1987ApJ,Kahn1974AA,Larson1971AA} --- is that this accretion
occurs through a disk \citep{Kuiper2010ApJ,Yorke2002ApJ,Nakano1989ApJ}. 

The signposts of disk accretion toward embedded HMYSOs are mainly of two kinds: 
highly collimated (aperture $\le5^\circ$) fast jets with velocities comparable to the escape speed 
of the compact object, and compact (sub-1000 au) circumstellar disks.
Slower, less collimated outflows can be produced by other processes like
ionization feedback \citep{Peters2012ApJ}, magnetic braking \citep{Hennebelle2011AA}, 
and mergers of smaller stars \citep{Bally2005AJ}. Rotating structures $\sim$3000 au or bigger
\citep{Beltran2016AARv} are too large to link them necessarily with disk accretion 
onto a central, compact  HMYSO, and may even be transient 
non-equilibrium structures (so-called toroids, \citealp{Cesaroni2006Nat}).

Examples of HMYSOs  in which collimated fast jets 
and compact sub-1000 au disks are detected simultaneously are 
GGD 27 \citep{Girart2017ApJ,Masque2015ApJ} and 
Cepheus A HW2 \citep{Torrelles2007ApJ,Curiel2006ApJ}. In these two HMYSOs, 
a compact disk is detected in molecular lines and the proper motion of the ionized lobes excited by the 
underlying jet indicate velocities in excess of 500 \kms. Additionally, 
 \citet{Jimenez-Serra2011ApJ} determined in Cepheus A HW2 the possible presence of high-velocity ionized gas associated with the jet 
by measuring the width of the associated  hydrogen  recombination lines (HRLs).
Sources G343.1262$-$0.0620 (also IRAS16547$-$4247, \citealp{Zapata2019ApJ,Rodriguez2008AJ}) 
 and G35.20$-$0.74N \citep{Beltran2016AA,Sanchez-Monge2013AA}, 
are also good examples, although the presence of a binary companion somewhat  confuses the proper motion signals 
from the ionized lobes.  In all these sources, the determined velocities are in excess of 200 \kms, which is the approximate 
escape velocity of a 8 \Msun\ star of  radius 100 $R_\sun$. 
This large radius is expected during the early phase of the HMYSO due 
to the large accretion rates \citep{Zhang2014ApJ}, and 200 \kms\ is a rough lower limit on the expected velocities associated with the gas
tracing accretion onto the central object.
Examples such as IRAS 13481$-$6124 \citep{Caratti-o-Garatti2016AA,Boley2016AA} and
W33A VLA1 \citep{Davies2010MNRAS} are less embedded, 
which allows its study in the infrared (IR).  Disks in these sources
are detected through near- and mid-IR interferometry (IRAS 13481$-$6124) and CO overtone mapping (W33A). 
The  high-velocity ionized gas in these sources
traces bipolar geometries as determined using Br$\gamma$ spectro-astrometry. 
In addition, W33A is associated with an ionized jet detected in cm free-free continuum \citep{Sanna2018AA}.
High-velocity ($\ge 200$ \kms) ionized gas is more commonly detected in the near-IR through (inverse) P-Cygni profiles of
HRLs, which provides no constraints to  the geometry or orientation of the outflow. 
One noticeable example, G018.3412+01.7681 (IRAS 18151$-$1208 B), is associated with an ionized jet \citep{Rosero2019ApJ},
a perpendicularly rotating  toroid   \citep{Beltran2016AARv}, 
and a Br$\gamma$  line displaying a P-Cygni profile \citep{Cooper2013MNRAS}.
\citeauthor{Cooper2013MNRAS} also present another 20 P-Cygni HMYSOs 
detected in spectra with 600 \kms\ spectral resolution (suggesting even larger expansion velocities).
G094.6028$-$01.7966 is another interesting source in which the Br$\gamma$ P-Cygni
signature is present \citep{Pomohaci2017MNRAS} and is associated with 
non-thermal jet lobes \citep{Obonyo2019MNRAS}. 
Eight more HMYSOs associated with P-Cygni 
profiles are shown in \citet{Pomohaci2017MNRAS}.
Finally, we note  S255 SMA 1 (also S255 NIRS 3), associated with a molecular disk, a jet,   
\citep{Cesaroni2018AA,Zinchenko2015ApJ}, 
and a high velocity Pa$\beta$ P-Cygni profile reported as a private communication in \citet{Cesaroni2018AA}.

Perhaps not surprisingly, in most cases the high-velocity gas is ionized \citep[with few exceptions, cf.][]{Titmarsh2013ApJ}. 
Because of the small distances to the HMYSO and the fast shock velocities involved, molecules 
are likely dissociated. In general, velocities probed by water masers are the highest among those probed by the other  molecular outflow tracers.
Examples of disk+outflow systems detected in molecular maser transitions, with associated ionized jets, are given 
in Table 3 of \citet[][\chg{that includes the noticeable example IRAS 20126+4104, e.g., \citealp{Chen2016ApJ}}]{Sanna2018AA}, to which we add 
G16.59$-$0.05 \citep{Moscadelli2019AA,Moscadelli2013AA558},
G353.273+0.641 \citep{Motogi2013MNRAS,Motogi2019ApJ}, 
 likely W75N(B) VLA 2 \citep{Carrasco-Gonzalez2015Sci}, 
 and possibly NGC 7538 IRS 1 (although results in \citealp{Beuther2017AA} favor a different interpretation).
Finally, pseudo-thermal (as opposed to strong masers) molecular lines have revealed several examples of compact disks and 
collimated molecular outflows associated with HMYSOs.  Among these, noticeable examples are those detected toward
AFGL 4176 \citep{Johnston2015ApJ}, 
G11.92$-$0.61 MM1 \citep{Ilee2018ApJ,Ilee2016MNRAS}
\gds\ \citep[also CRL 2136][]{Maud2019AA,Maud2018AA}, 
Orion Src I \citep{Ginsburg2018ApJ,Hirota2017NatAs}, G343.1262$-$0.0620 \citep{Zapata2015MNRAS,Zapata2019ApJ}


All these examples demonstrate that disk accretion plays a role in assembling  massive stars,
although, what fraction of the mass is accreted  in this way is still unclear.
Theoretically, disk accretion should be effective after the protostar has contracted and it should allow accretion 
to proceed until reaching the star's  final mass \citep{Sartorio2019MNRAS,Kuiper2018AA,Keto2007ApJ}, 
in spite of the presence of UV radiation and  ionized gas.  
At least for accretion rates $\le10^{-3}$ \Msun\ yr$^{-1}$ \citep{Hosokawa2010ApJ}, 
the short Kelvin-Helmholtz timescale in HMYSOs   ensures that a large fraction 
of the mass in O-type stars will be accreted after protostellar contraction.
In the simulations by \citet{Zhang2014ApJ}, for example, they show that an accreting  HMYSO reaches  
surface temperatures $\ge20000$ K --- sufficient to excite an \hii\ region --- when it has accreted $\sim12$ \Msun, that is, 
about half of its final mass of 26 \Msun.

However, most of the HMYSOs above  
have no
associated \hii\ region. 
An ALMA search for compact disks
toward HMYSOs with signs of rotation at $>10^3$ au scales performed by  \citet{Cesaroni2017AA}  suggests 
that more evolved targets --- that is, those associated with hyper-compact (HC) \hii\ regions ---
seem to lack a compact molecular disk counterpart when observed with high angular resolution 
  (at least in CH$_3$CN). 
While the specific reason for disk non-detection may be somewhat related to the selected tracer 
(e.g.\ G17.64+0.16, \citealp{Maud2018AA}), 
there are alternative pictures in the literature which do not  necessarily require a 
long-lived disk from where the HMYSO gathers its mass after the HC \hii\ region 
stage \citep{Goddi2018arXiv,DePree2014ApJ}.

In this work, we focus on G345.4938+01.4677 (G345.49+1.47 hereafter), 
a $L_{\rm bol}=1.5\times10^5 L_\sun$ HMYSO located at 2.38 kpc \citep{Lumsden2013ApJ,Mottram2011AA,Moises2011MNRAS}. 
\hmyso\ is associated with an HC \hii\ region, a massive molecular outflow \citep{Guzman2011ApJ},
a high-velocity ($\gtrsim600$ \kms) ionized jet \citep{Guzman2016ApJ},
and a perpendicularly rotating core \citep{Guzman2014ApJ} of diameter $\sim4000$ au 
detected in sulfur oxide  lines (SO, SO$_2$ and isotopologues). 
G345.49+1.47 is part of a small group of HMYSOs toward which active disk accretion has been proposed 
while simultaneously exciting an \hii\ region. Other like sources are G35.20$-$0.74N \citep{Beltran2016AA} and
recently W75N (B) VLA1 \citep{Rodriguez-Kamenetzky2020MNRAS}.
G345.49+1.47 was  also part of the sources studied by \citet{Cesaroni2017AA} at 0\farcs2 angular resolution. 

Here, we present and analyze new high-angular resolution ($\approx22$ mas, equivalent to 52 au) 
line (\hrl\ HRL) observations of \hmyso\ and its adjacent continuum. 
High angular resolution HRL observations have been used to trace the dynamics (rotation and  outflows) of ionized gas
around HMYSOs such as IRAS 07299$-$1651 (also G232.6207+00.9959, \citealp{Zhang2018NatAs}),
G45.47+0.05 \citep{Zhang2019ApJ},
and \gds\ \citep{Maud2019AA}.
Section \ref{sec:obs} presents the observations and describes the data reduction. 
Observational results are presented in Section \ref{sec:res} and the main analysis and discussion in Section \ref{sec:dis}.
The main conclusions of this work are summarized in Section \ref{sec:con}. 

{\section{Observations.}\label{sec:obs}}

Observations of \hmyso\ were taken using the ALMA Band 6 in two sessions taken the 8th and 9th of July 2019 (UT). 
They consisted of 12m-array single pointing (field of view FWHM $27\arcsec$)
scans toward R.A.=16$^{\rm h}$59$^{\rm m}$41\fs63, decl.=$-40\arcdeg03\arcmin43\farcs5$ (ICRS) 
in five spectral windows (SpWs) centered \chg{at the frequencies of} 
219.54, 220.41, 220.43, 222.75, and 224.72 GHz 
and covering  1875, 234.4, 234.4,  1875, and 234.4 MHz, respectively.
The bandpass and primary flux calibrators for the 8th and 9th of July sessions were 
J1924$-$2914 and  J1427$-4206$, whose fluxes at  222.75 GHz were fixed at 2.817 and 1.077 Jy, respectively. 
The absolute flux calibration of ALMA in band 6 is estimated to be accurate within 10\% \citep{ATH7}.

Table \ref{tab:ALMA_obs} shows some 
observational parameters of the data: columns (1) to (9) show 
the ALMA band, the number of antennae employed, the reference frequency, 
the  flux of the phase calibrator J1711$-$3744 at the reference frequency, 
the date of the observations, the on-source time, 
the approximate  maximum angular resolution (MAR, given by $\lambda/L_{\rm max}$, where $L_{\rm max}$ is the largest baseline), 
the maximum recoverable scale (MRS, given by $0.6\lambda/L_{\rm min}$, where $L_{\rm min}$ is the shortest baseline, see Eq.\  (3.28) in \citealp{ATH7}),
and the median $T_{\rm sys}$ measured during the observations.

We focus our analysis on the 1875 MHz wide SpW centered at 222.75 GHz ($\sim1.3$ mm).
This SpW covers the \hrl\ HRL whose rest frequency is  222011.75545 MHz. The
velocity width of the channels covering the \hrl\ line is 1.32 \kms, and the spectral resolution of the data cubes is 2.64 \kms.

Calibration and reduction of these data were done using the 
\emph{Common Astronomy and Software Applications} \citep[CASA,][]{McMullin2007ASPC} v.\ 5.4. 
A priori calibration based on external calibrators was carried out by the ALMA East-Asia regional center. 
A continuum level was determined from the  calibrated uv-data from the line-free channels of the SpW and 
 subtracted  using the task \texttt{uvcontsub}.
Continuum images of the 1875 MHz SpW were  obtained with  the task \texttt{tclean} using multi-frequency synthesis and
 Briggs weighting  \citep{Briggs1995PhD} using robust parameter set to 0. 
 Several iterations of phase self-calibration were
 applied to the continuum, until reaching a gain calibration solution sampled every 15 s. 
 These gain tables were later applied to the continuum subtracted uv-data, from where 
 image cubes were synthesized following the same procedure and uv-weighting scheme as 
 with the continuum, but channel per channel. The total mapped region in both continuum and 
 line was selected to cover the region where the primary beam response is above 0.2, corresponding to a 
diameter of $\approx39\arcsec$ around the phase center.
Table \ref{tab:ALMA_ima} details some observational characteristics of the final reduced data products.
Calibrated images and cubes of the HRL can be found  in \citet{Guzman2020Dataset_G345disk}.
 
\begin{deluxetable}{llllllllc}
\tablecaption{ALMA observational parameters.\label{tab:ALMA_obs}}
\tablecolumns{9}
\tablehead{
	\colhead{Band} &\#& \colhead{Reference}& \colhead{Phase\ cal. }& \colhead{Observation }& \colhead{${t_{\rm on}}^\ddagger$}& \colhead{MAR$^\star$  }&\colhead{MRS$^\dagger$}  & \colhead{$T_{\rm sys}$}\\ 
                    	&\colhead{Ant.}& \colhead{frequency} & \colhead{flux} & \colhead{date} &  &  & & \\
            &&  \colhead{(GHz)}&  \colhead{(mJy)} &  \colhead{(dd-mm-yyyy)} & \colhead{(s)}& \colhead{(mas)} &  \colhead{(\arcsec)} & \colhead{(K)}
            }
\startdata
6  & 46 & 222.747 &116.5 & 08-07-2019 & 2624 & 18.5 & 1.12 & 85   \\
6  & 44 & 222.747 &111.4 & 09-07-2017 & 2622 & 19.9 & 1.12 & 75    \\
\hline
\multicolumn{9}{l}{$^\ddagger$ Time spent on the science source as reported in the observing log.} \\
\multicolumn{9}{l}{$^\star$ Maximum angular resolution, given by  $\lambda/L_{\rm max}$.} \\
\multicolumn{9}{l}{$^\dagger$ Maximum recoverable scale, given by $0.6\lambda/L_{\rm min}$ (Eq.\  (3.28) in \citealp{ATH7}).} \\ 
\enddata
\end{deluxetable}

\begin{deluxetable}{lllcc}
\tablecaption{Observational characteristics of obtained images\label{tab:ALMA_ima}}
\tablecolumns{5}
\tablehead{
\colhead{Image} & \colhead{Spectral} & \colhead{Synth.\ beam} & \colhead{rms} & \colhead{Dynamic}\\
\colhead{} & \colhead{resolution} & \colhead{bmaj$\times$bmin, pa} & \colhead{} & \colhead{range}\\
\colhead{} & \colhead{(\kms)} & \colhead{(mas\,$\times$\,mas, $^\circ$)} & \colhead{($\mu$Jy beam$^{-1}$)}&\colhead{}
}
\startdata
Continuum & \nodata & $24\times20$, $81.4$ &  30 &  2300 \\
\hrl\ cube    &  2.64    & $24\times20$, $82.0$  & 696  &     12\\
\enddata
\end{deluxetable}

{\section{Results}\label{sec:res}}

\begin{figure}
\hspace*{-1ex}\includegraphics[width=0.6\textwidth]{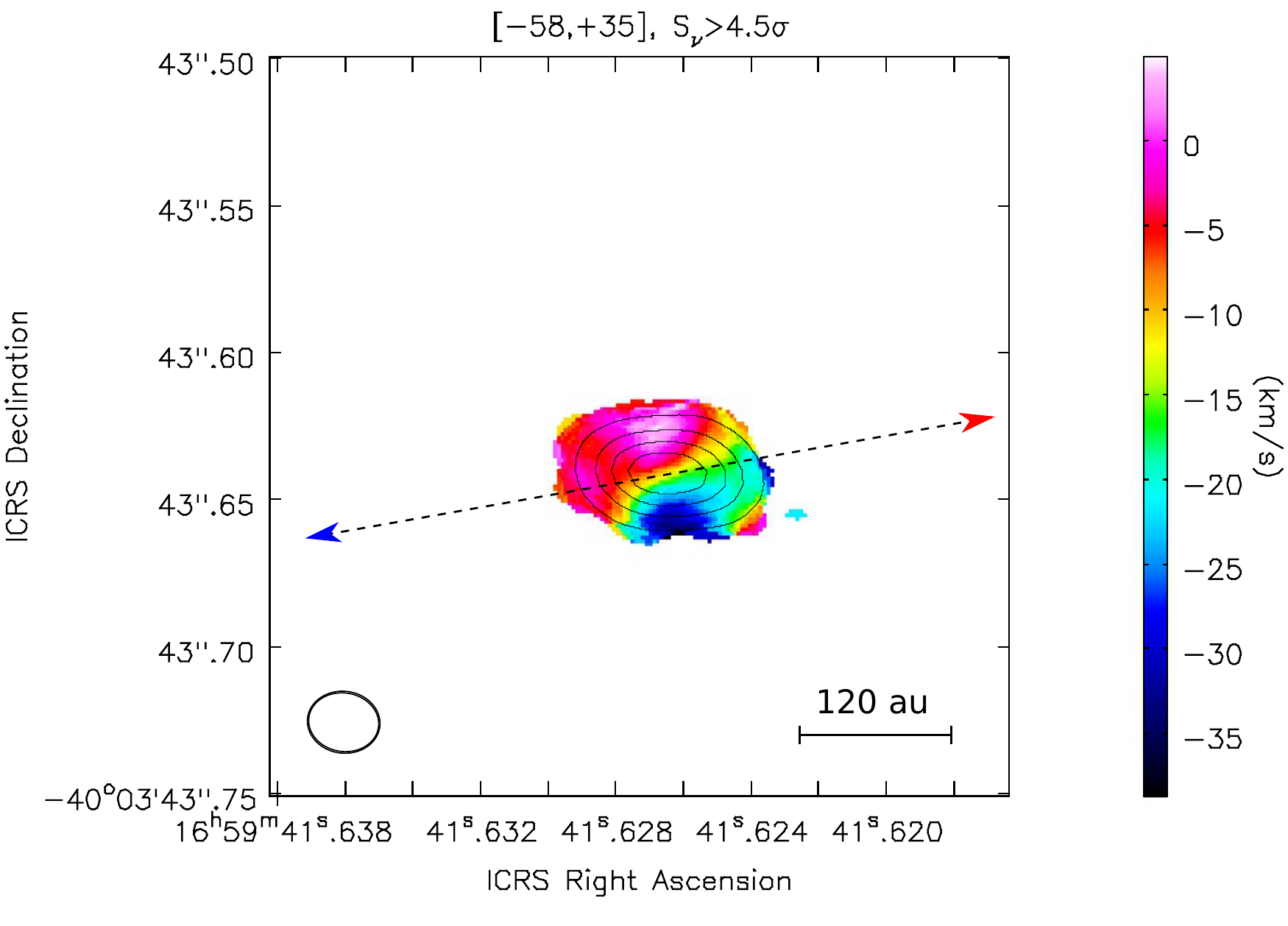}~~%
\includegraphics[width=0.4\textwidth]{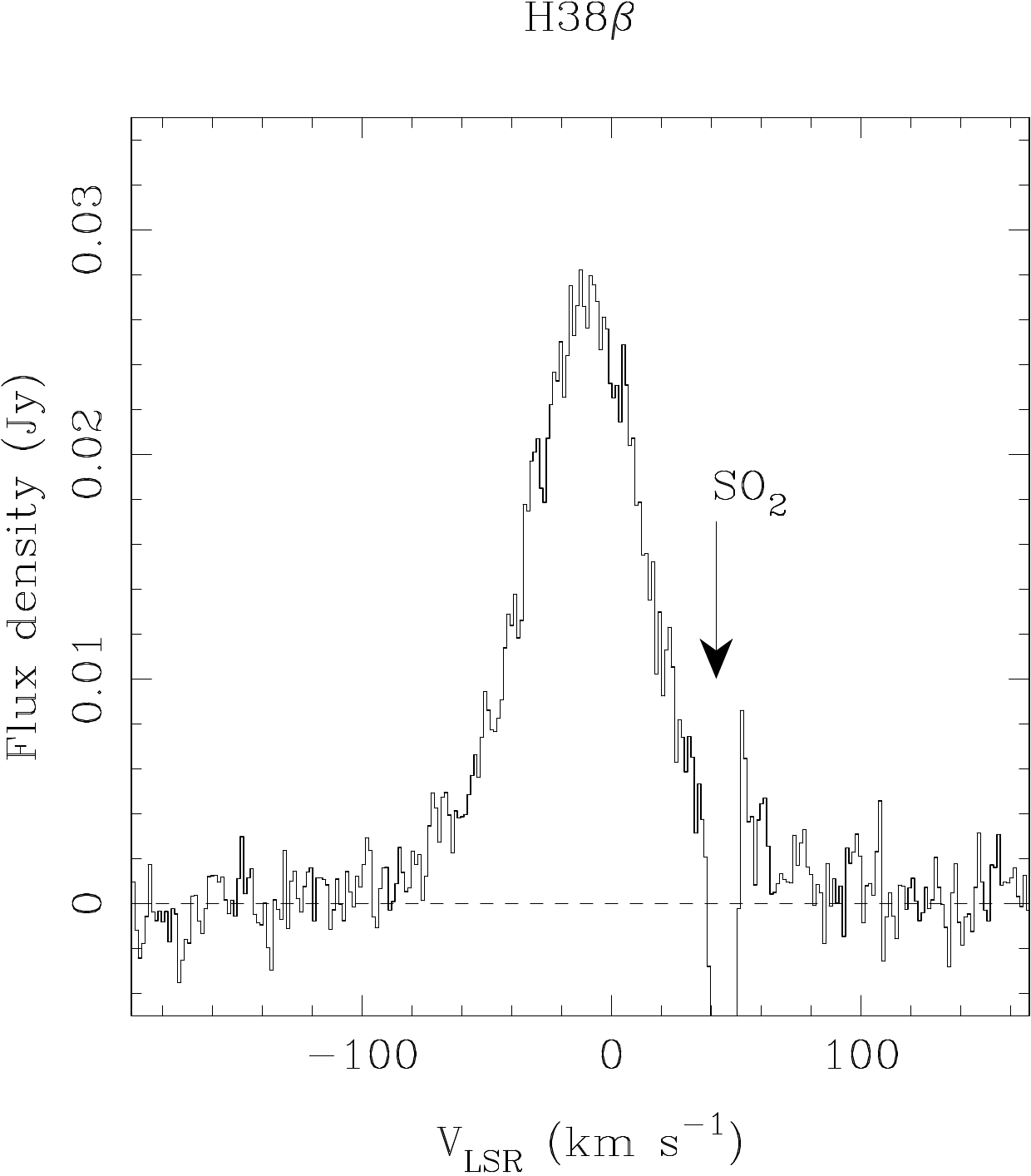}%
\caption{Left panel: moment 1 of the \hrl\ line detected toward \hmyso, with black contours showing the continuum. 
Contour levels are 20, 40, 60, and 80\% of the peak (69.1 mJy beam$^{-1}$). 
Moment 1 was calculated using all emission over 4.5 $\sigma$ ($\sigma=0.696$ mJy beam$^{-1}$) 
and $v_{\rm LSR}$ between $-58$ and $+35$ \kms.  The dashed double arrow shows 
the P.A.$=100^\circ$ direction of the protostellar jet \citep{Guzman2016ApJ}. The scale bar shown in the bottom right 
assumes a distance of 2.38 kpc. 
Right panel: Flux density of \hrl\ calculated within a circle of 50 mas diameter centered at the
peak of the  continuum position. \label{fig:cont_m1_profile}}
\end{figure}

Figure \ref{fig:cont_m1_profile} shows the continuum and \hrl\ data obtained toward \hmyso. Left panel shows the  \hrl\ moment 1 
overlaid with continuum levels. As shown by the contours, 
the spatial distribution of the resolved continuum emission at 222.7617 GHz near \hmyso\ matches the 
distribution of the line, at least over the first contour at $14$ mJy beam$^{-1}$.
This indicates that a  large fraction of this continuum emission comes from  free-free. 
We fit a 2D Gaussian to the continuum emission using the task \texttt{imfit} within CASA. 
Table \ref{tab:contsou} shows the best-fit  source parameters. 
Throughout this work we will use the  position of this source as reference whenever offsets are used as spatial axes.

\begin{figure}
\includegraphics[width=\textwidth]{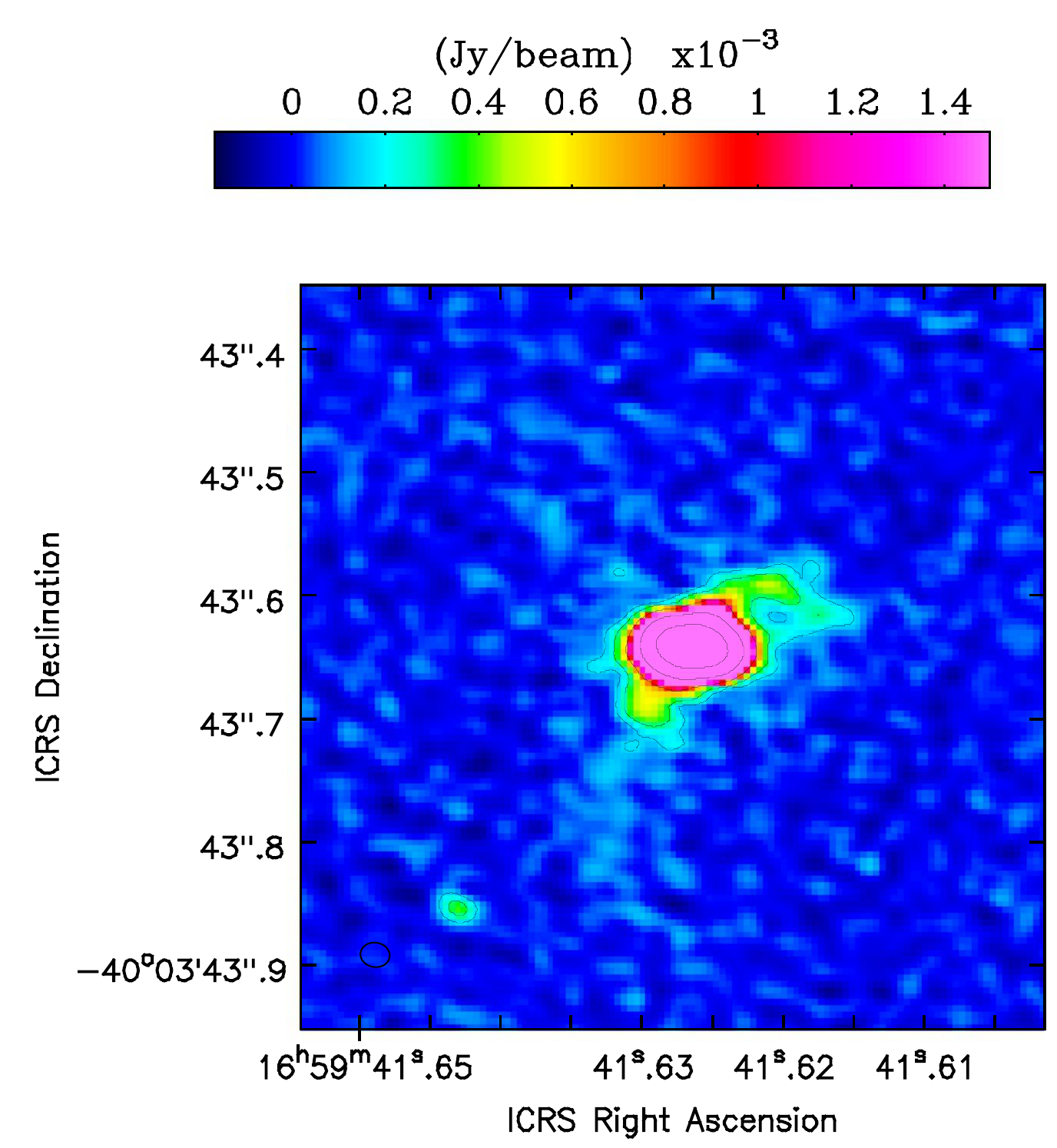}~~%
\caption{Continuum emission within $0\farcs6\times0\farcs6$ centered in \hmyso. The color scale stretch has been saturated to 
emphasize low-brightness features. Contour levels are $-5$, 5, and $5+5^i\times\sigma$ with $i=1\ldots4$ and $\sigma=30\,\mu\text{Jy}$ beam$^{-1}$. \label{fig:contsat}}
\end{figure}

Figure \ref{fig:contsat} shows the low-brightness continuum emission around \hmyso. 
We detect  emission surrounding the HMYSO  protruding in 
 the  north-west and south-east directions extending 0\farcs1, although there is a hint of 
 additional   filamentary emission extending toward the south for 0\farcs25.
We  integrate the intensity within 0\farcs3 around \hmyso\  and report it in Table \ref{tab:contsou}.
While the extended  features detected  are within the MRS, 
we emphasize that (by definition) the recovered brightness fraction of 
a structure homogeneous within the  MRS size is approximately equal to  $1/e$ or $37\%$.
Hence, part of the emission of the more extended features in  Figure \ref{fig:contsat} likely are being partially filtered.

In addition, we report the detection of unresolved Source 10a (\hmyso\ is Source 10 in \citealt{Guzman2014ApJ})
 located 0\farcs29  to the south-east  from \hmyso, or $\sim690$ au in projection. The flux density 
 of this source is smaller than the rms in the data presented by \citet{Cesaroni2017AA}, but we are able to detect it with 
 a SNR $>6.5$. 

To evaluate whether Source 10a may be part of extragalactic contamination, we extrapolate with a power-law the  source counts from the survey of \citet{Mocanu2013ApJ} at 220 GHz, obtaining a source density of $24/S_{\rm mJy}^{2.3}$ deg$^{-2}$ mJy$^{-1}$. Based on this extrapolation, we estimate  $\sim63$ extragalactic sources per deg$^2$ as bright as 10a, 
or $\lesssim6\times10^{-3}$   sources within the primary beam of our observations.
 We conclude that Source 10a is Galactic in nature and part of the IRAS 16562$-$3959 clump.

%
\begin{deluxetable}{llllll}
\tablecaption{Continuum emission from sources  in \hmyso\ field at 222.76 GHz  \label{tab:contsou}}
\tablecolumns{6}
\tablehead{
\colhead{Source} 	& \colhead{R.A.}	& \colhead{Decl.} 	& \colhead{Flux} 	& \colhead{Deconvolved size}		& \colhead{Integrated}\\
\colhead{} 		& \colhead{ICRS} 	& \colhead{(ICRS)} 	& \colhead{density} 	& \colhead{bmaj $\times$ bmin, P.A.}&\colhead{flux density\tablenotemark{{\rm \dag}}}\\
\colhead{} & \colhead{(16$^{\rm h}$:59$^{\rm m}$:)} & \colhead{($-40$\arcdeg:03\arcmin:)} & \colhead{(mJy)}&\colhead{(mas $\times$ mas, \arcdeg)} & \colhead{(mJy)}
}
\startdata
\hmyso 	& 41\fs6264 & 43\farcs641 &  $165.6\pm1$ 	&  $35.93\pm0.3\times16.32\pm0.2$, $86.8\pm1$& $173\pm1$ \\
Source 10a    	& 41\fs6430 & 43\farcs853 &  $0.403\pm0.06$ 	& { unresolved}							& $0.39\pm0.1$ \\
\enddata
\tablenotetext{\dag}{Within 0\farcs3 for \hmyso\ and within 50 mas for 10a.}
\end{deluxetable}

\begin{figure}
\includegraphics[width=\textwidth]{{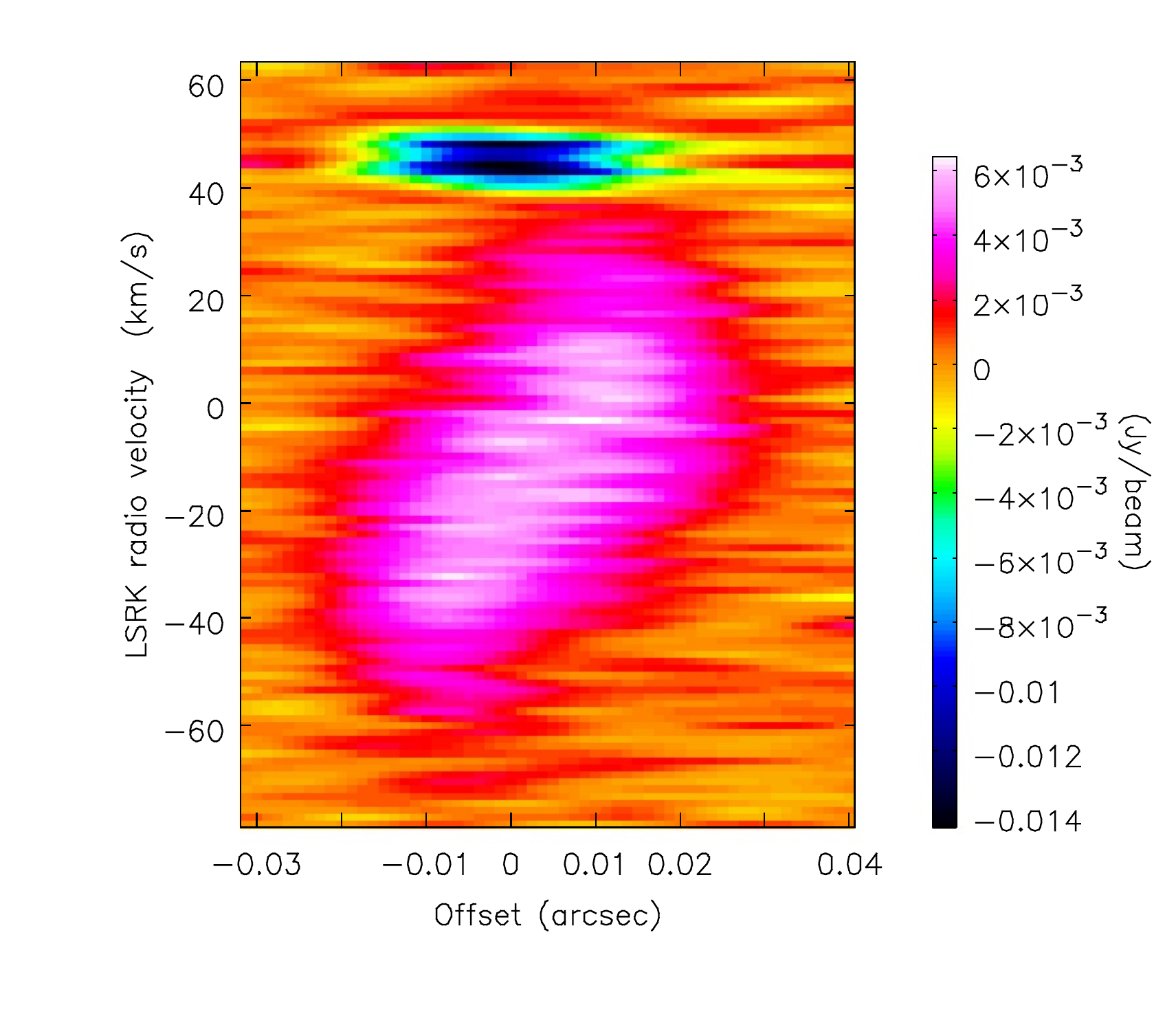}}%
\caption{Position velocity diagram of \hrl\ (rest frequency) centered in \hmyso\ taken along the P.A.$=10\arcdeg$ direction from the south-west end (negative offset)
to the north-east. We averaged one beam width across the position velocity slice. The SO$_2$ absorption is evident in the red-shifted part of the diagram. \label{fig:pv}}
\end{figure}

The right panel of Figure \ref{fig:cont_m1_profile} shows the  \hrl\  spectrum integrated 
in a circle of 50 mas around \hmyso. 
The HRL is characterized by a width of $58.4\pm1$ \kms, 
is centered around  $-11.0\pm1$ \kms,  and \chg{has} a 
total flux of $W_L=1.629\pm0.03$ Jy \kms. 
\chg{The noise level of this integrated  spectrum is $1.4$ mJy.}
The line flux peaks at  $8.12$ mJy beam$^{-1}$  ($T_B\approx420$ K).  
The molecular line absorption shown in Figure \ref{fig:cont_m1_profile}
is due to the  SO$_2\,(v=0)$  $J_{K_a,K_c}=11_{1,11}\rightarrow10_{0,10}$ line centered at $221965.22$ MHz. This line has
an upper energy level of $E_{\rm u}/k_B=60.4$ K \citep{Pickett1998JQSRT}. We observe 
emission from this line on a more extended spatial distribution than that of  the HRL.
The \hrl\ moment 1 map in Figure \ref{fig:cont_m1_profile} clearly shows we are able to resolve the \hrl\ emission. 
This is further illustrated by the position velocity diagram (pv-diagram) of the HRL shown in Figure \ref{fig:pv}.
\chg{The position velocity diagram shows the average intensity over one beam width across a P.A.$=10\arcdeg$ line centered on the position of \hmyso\ and 
has a noise level of 0.51 mJy beam$^{-1}$.}
The $v_{\rm LSR}$ range used to calculate the moment 1 is \chg{$-58\le v_{\rm LSR}(\text{km s}^{-1})\le+35$}, \chg{including only emission above $4.5\sigma$ and excluding the line wings.  
This threshold allows us to negate the influence of the SO$_2$ absorption affecting the red-shifted slope of the line. The 
 statistical uncertainty derived of the velocity map ranges between $2.7$--$3.8$ \kms, depending somewhat on the number of channels included
 in the moment 1 calculation for each position. 
Figure \ref{fig:pv} illustrates that most of the HRL emission is not affected by the molecular absorption.}
We measure a velocity gradient in the P.A.$=11\arcdeg$ direction of $+1.13$ \kms\ mas$^{-1}$ ($+0.47$ \kms\ au$^{-1}$ at 2.38 kpc).
\hrl\  is not detected toward any other location of the map. The pv-diagram of the HRL 
somewhat resembles that of H30$\alpha$ in \gds\ \citep[][their Figure A.1]{Maud2019AA}.

\FloatBarrier{\section{Discussion}\label{sec:dis}}

In this section we analyze in more detail the physical interpretation of the HRL and its characteristics. One important physical parameter 
for this analysis is the distance to \hmyso. For this work, we use the spectrophotometric 
distance of 2.38 kpc  \citep{Moises2011MNRAS} instead
of the kinematic distance of 1.7 kpc used by several authors before. Both estimations have appeared in the literature.
However, results from \citet{GAIA2018AA,GAIA2016AA} make the 2.38 kpc distance estimation more plausible \chg{taking into account}
the  low extinction levels $A_G=2.84$ and 1.82 \citep{Andrae2018AA}
and the parallax  distance estimations of $1983^{+490}_{-330}$ and $2020^{+200}_{-160}$ pc \citep{Luri2018AA} determined toward
2MASS sources J16594297$-$4003114 and J16593928$-$4004080, respectively. 
These two sources are in the field of the massive molecular clump IRAS 16562$-$3959,  of which \hmyso\ is the central, dominating HMYSO. 
However, from the 1.2 mm dust emission and the molecular clump model of \citet{Guzman2010ApJ}, we derive an
 H$_2$ column density in the direction of these two sources of about $4\times10^{22}$ cm$^{-2}$, which is 
 more than an order of magnitude larger than those  derived from the extinction values, suggesting that these two near-IR sources
 are in the foreground of IRAS 16562$-$3959. The appearance of these sources in three-color JHK$_S$ 
images also suggests they are in the foreground, situating \hmyso\ farther than 2 kpc.

{\subsection{HRL and Continuum Emission}\label{sec:basic}}

Because both the free-free emission and the HRL  arise from the same ionized gas, 
we can use the \hrl\ line to estimate the free-free contribution to the observed continuum emission. 
Under LTE conditions, the \hrl\ optical depth 
is given by $ \mathcal{T}_{L}\phi(\nu)$, 
where $\mathcal{T}_L$ is defined in  \citet[][Eq. (B5)]{Guzman2014ApJ}  and 
$\phi(\nu)$ is the line profile. 
We define the
line to continuum equivalent width $\Delta v_L:=\mathcal{T}_L/\tau_{\rm ff}$, where  $\tau_{\rm ff}$ is the free-free opacity. For the 
 the \hrl\ recombination line  we have 
\begin{equation}
 \Delta v_{38\beta} = 21.852\,\text{km~s}^{-1}\left(\frac{T_e}{10^4~{\rm K}}\right)^{-1.3}~~,\label{eq:equiw}
\end{equation}
where the dependance on the temperature is adequate for the frequency of this HRL. 
From \eqref{eq:equiw}, we can derive the LTE ratio between the  integrated line flux to the  free-free continuum. That is, 
\begin{equation}
\frac{W_{38\beta}}{S_{\rm ff}} = \Delta v_{38\beta}\, \mathcal{C}(\tau_{\rm ff})~~,\label{eq:l2c}
\end{equation}
where $\mathcal{C}(\tau_{\rm ff})\le1$ is an opacity  correction factor which is unity for optically thin emission and decreases to zero
monotonically with increasing $\tau_{\rm ff}$.
We argue that LTE should be a reasonable approximation in \hmyso\ because the electron density is high enough to
ensure  thermalization via collisions.  
The  thermalization critical density is $\sim9\times10^6$ cm$^{-3}$ \citep[][]{Strelnitski1996ApJ} for \hrl, which is lower than the 
electron density found by  \citet{Guzman2014ApJ} at even  larger physical scales than the ones probed by this study.
Hereafter, we will use $T_e=7000$ K which is adequate for \hii\ regions located at the Galactocentric distance of 
\hmyso\ \citep{Paladini2004MNRAS}. 

Observationally, the  integrated line flux to continuum ratio in \hmyso\ is $9.84\pm0.2$ \kms, which is lower than the value of $13.74$ 
\kms\ obtained from  Equation \eqref{eq:equiw} using $7000$ K.
%
Because the continuum emission does not seem to follow the power-law --- characteristic of partially optically thick free-free emission ---
determined in \citet{Guzman2016ApJ}, it is 
likely that the free-free emission at 220 GHz from the HC \hii\ region is already optically thin.
Therefore, we propose that the excess continuum emission \chg{relative to that inferred from}  the HRL arises from thermal dust emission.
This dust emission contribution \chg{is} about $119$ mJy. Considering that the dust emission estimated by \citet{Guzman2014ApJ} at 92 GHz is $\le11$ mJy, the derived 220 GHz dust emission is consistent with the observed ranges of dust  mm spectral indices \citep[e.g.,][]{Orozco2017ApJ}.

{\subsection{A Photoionized Disk}\label{sec:disk}}

We  interpret the velocity profile of the \hrl\ line as rotation around the HMYSO \hmyso. 
In this case, a simple model for the emission is that of a \chg{Keplerian disk 
as used in G17.64+0.16} by
\citet{Maud2018AA}. 
\chg{However, we must first note that}, when considering only rotation, there is a diameter of the disk whose velocity projected in the line of sight (\los) 
is zero. 
This diameter corresponds to the minor axis of the ellipse  projected by the disk in the plane of the sky.
That is, emission 
at the source's $v_{\rm LSR}$   traces a straight line in the velocity moment 1 map.  However, 
as shown in  Figure  \ref{fig:cont_m1_profile},  the emission at the source $v_{\rm LSR}$ --- which separates the blue- and red-shifted 
halves of the rotating structure --- clearly  traces a slanted serpentine shape, to which we will refer hereafter as a \cs-shape.\footnote{Alternatively, the moment 1 panels shown in Figure  \ref{fig:cont_m1_profile} can be described as resembling a \emph{taijitu} or \emph{ying-yang} symbol.} This \cs-shape is characteristic of radial motions combined with rotation, specifically, 
radial motions in which the rotation to radial velocity ratio is not constant with radii. 
Therefore, in order to reproduce this observed feature, we include 
 radial motions in the disk in addition to rotation.
Compared to the pv-diagram analysis of SiO in G17.64+0.16 done by \citet{Maud2018AA}, our  \hmyso\ pv-diagram (Figure \ref{fig:pv}) 
seems  to be less informative, possibly due to HRLs being much wider than molecular lines.
Therefore, in our case, the main indication of radial motions relies on the aspect of the moment 1 map rather than that of the pv-diagram.

The model consists of a thin circular disk of ionized gas  whose position, orientation, and $v_{\rm LSR}$ define five free parameters. 
The rotation is assumed to be Keplerian defined by a central mass, and the radial motion has a constant magnitude. 
The disk is characterized by a radial  electron emission measure \citep[EM, Equation (10.34) in ][]{ToRA6} distribution, \chg{truncated at an
 external radius $R_e$.}
 The disk also has a width $w_d$ which, in combination with the emission measure, 
 determines the electron density of the disk as 
 \begin{equation}
 n_e= \sqrt{\frac{{\rm EM}(r)}{w_d}}~~.\label{eq:ne}
 \end{equation} 
Electron density affects HRLs through pressure or collisional broadening: in general, the shape of an HRL is a Voigt profile, defined as the 
convolution of a Gaussian with a thermal width  and a 
Lorentz profile with width parameter given by \citet[][Equation (2.74)]{Gordon2009ASSL}. For \hrl, the Lorentz width parameter is 
$8.75\,n_{e,8}$ \kms, where $n_{e,8}$ is the electron density in units of $10^8$ cm$^{-3}$.

\begin{figure}
\centering\includegraphics[height=0.55\textheight, trim={0  143mm 0 0},clip]{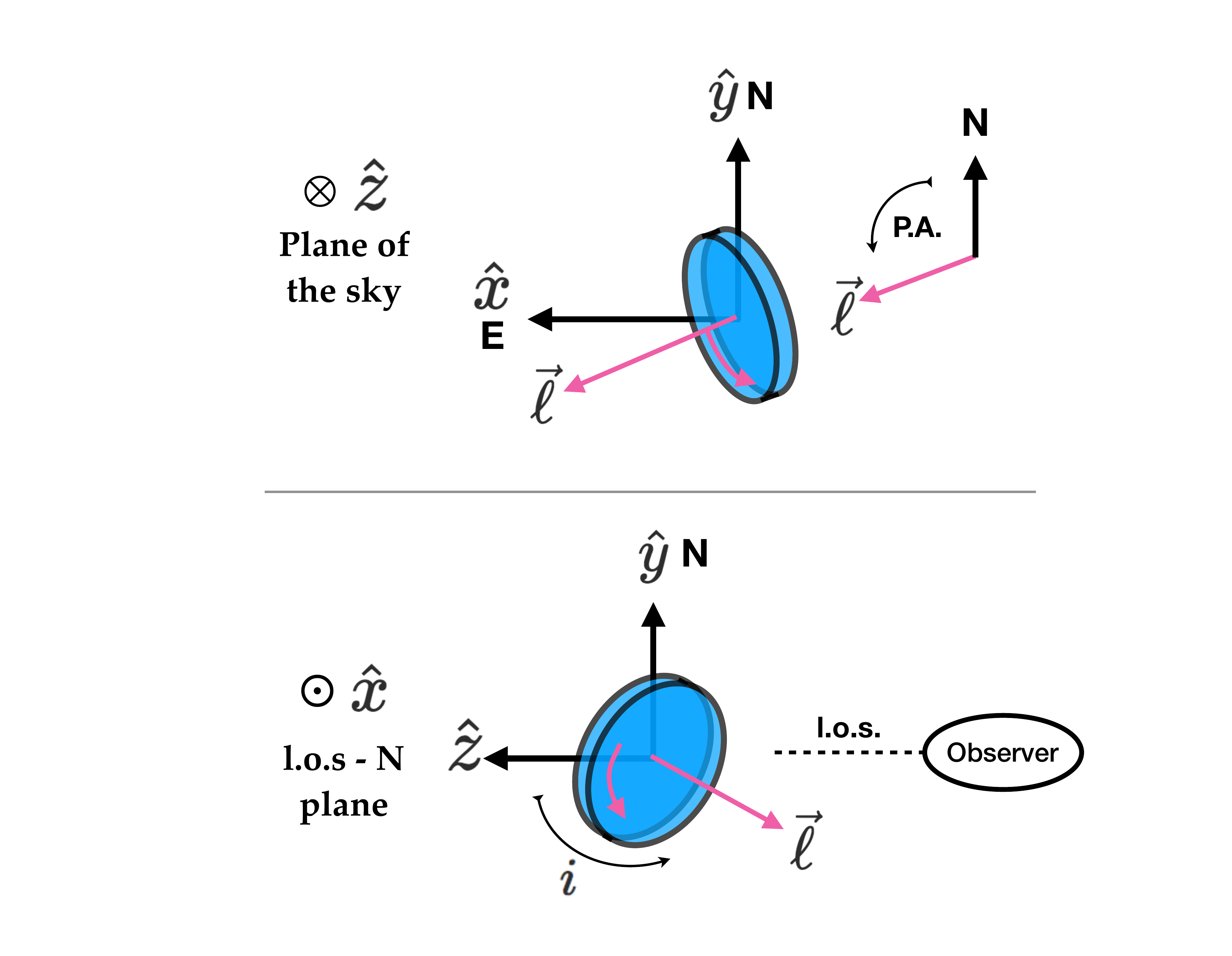}%
\caption{Coordinate system used to describe the \hmyso\ disk orientation. Top and bottom panels show
the projection of the system in the plane of the sky and the \los--north plane, respectively.  Note that 
angle $i$ is measured in the plane $\hat{z}$-$\vec{\ell}$ \cchg{(that is, the inclination of the disk axis respect to the plane of the sky is $i-90^\circ$)}. \label{fig:coords}}
\end{figure}

Before determining the best fit model, we define some constraints on the orientation of the disk based on previous studies. 
Let us consider a right hand oriented system of reference where 
$(\hat{x},\hat{y},\hat{z})$ directions are east, north, and the \los, respectively. The orientation and rotation direction of the disk is defined 
by the direction of its angular momentum $\vec{\ell}$. We define the  inclination angle $0\le i\le 180^\circ$ 
of the disk as the angle formed between $\hat{z}$ and $\vec{\ell}$. Figure \ref{fig:coords} shows schematically the projections of the 
disk orientation  in the adopted reference system. 
The P.A. and $i$ angle are measured in the plane of the sky and in the \los--$\vec{\ell}$ plane, respectively \cchg{(i.e., the inclination of the disk respect to the plane of the sky is $i-90^\circ$)}.
We will consider models in which the axis of the disk is consistent with the jet direction and rotation shown in Figure \ref{fig:cont_m1_profile}. 
This not only means  that P.A.$\approx100\arcdeg$, but also that the $\vec{\ell}$ direction 
corresponds to the  blue-shifted half of the jet which is closer to
the observer than the western half (Figure \ref{fig:coords}). This constraint, combined with the observed rotation,  
implies that $\vec{\ell}$ points toward us (that is, $\vec{\ell}\cdot\hat{z}<0$ \chg{or $i>90^\circ$}).

These geometric considerations solve the expansion-contraction degeneracy of the radial motions: for \hmyso,
 the \cs-shape noted above in the moment 1 map for velocities close to the 
$v_{\rm LSR}$ corresponds to \emph{expansion}. Contracting motions would create a $\csreflect$-shape pattern instead.

\begin{figure}
\includegraphics[width=\textwidth]{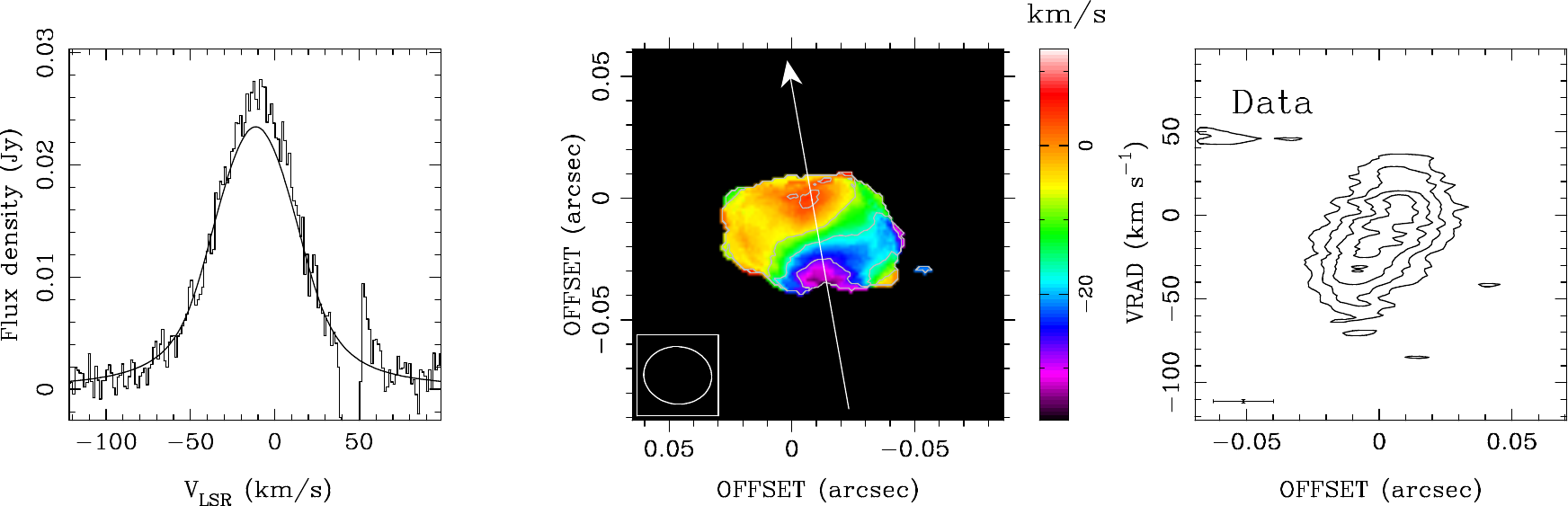}\\
\includegraphics[width=\textwidth]{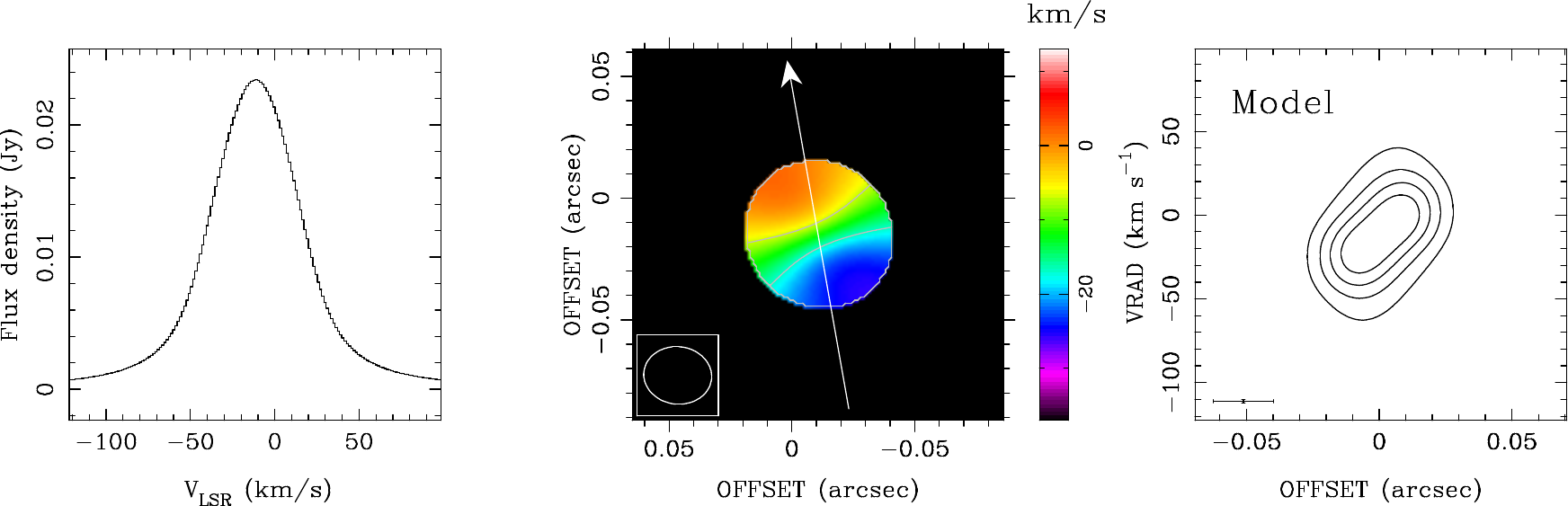}\\
\includegraphics[width=\textwidth]{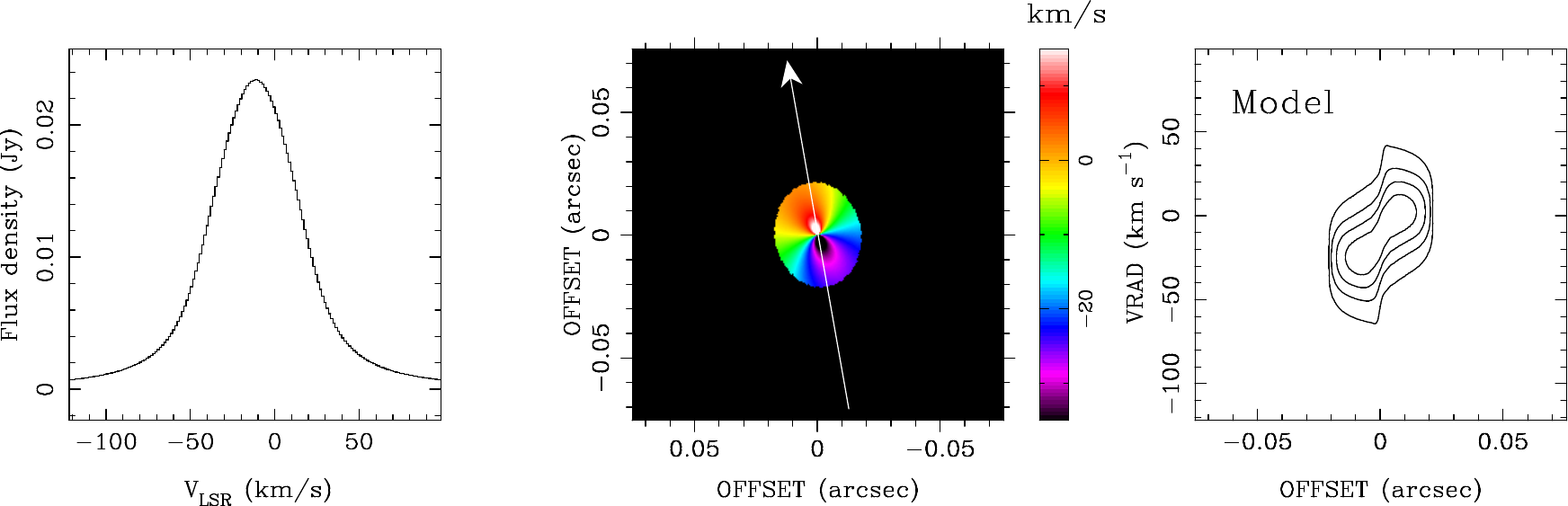}%
\caption{First row, from left to right: panels show the integrated spectrum, moment 1 (five contour levels between $-35$ and $15$ \kms),  
and pv-diagram (contour levels at 20, 40, 60, and 80\% of the  peak $=6.9$ mJy beam$^{-1}$), respectively. 
The arrow in the second panel shows the size and orientation of the pv-diagram cut. Velocities close to the 
$v_{\rm LSR}$ are colored in shades of green and trace the \cs-shape pattern evidence of radial motions.
Second row:  best-fit model (Table \ref{tab:fit}) convolved with the \chg{clean} beam. 
Model spectrum shown in the first  panel is the same as in the first panel of the first row. Contour levels 
in second and third plots are the same as the respective panels in the first row. 
Third row: theoretical best-fit model  not convolved with the \chg{clean} beam.  \label{fig:disk}}
\end{figure}

Figure \ref{fig:disk} shows the results of the model that best fits the data towards \hmyso.  
The leftmost column of panels  shows the total flux density spectra, the middle column
the moment 1 maps, and the rightmost column the position velocity diagrams. The top panel row shows these three 
diagnostic plots for the observed data. 
To determine the best-fit model, we calculate the 
simulated spectrum, moment 1, and pv-diagram and convolved them with the spatial resolution \citep{Richer1991MNRAS}. Note 
that the spectral resolution of the data is very small compared to the width of the line. 
The convolved models are shown in the second row of panels in Figure \ref{fig:disk}. The model shown in the first panel of this second row is the same as the one shown in the top-left panel.
The third row of plots also show  the three diagnostics (spectrum, moment 1, and pv-diagram) of the best-fit model, but not convolved with the beam. 
We calculate the difference between the convolved model and the data, and minimize the squared difference weighted by the inverse variance of the data. 
Table \ref{tab:fit} shows the best-fit parameters of the disk, together with those parameters which were held fixed.
The minimization and errors were calculated using the Minuit package \citep{MINUIT1975}, as implemented in the Perl Data Language. Computational routines can be found in  \citet{Guzman2020Dataset_G345disk}. 
Error bars in Table \ref{tab:fit} represent  statistical uncertainties.


\begin{deluxetable}{lll}
\tablecaption{Disk best-fit parameters. \label{tab:fit}}
\tablecolumns{3}
\tablehead{
\colhead{Parameter} 	& \colhead{Best-fit value} & \colhead{}
}
\startdata
$v_{\rm LSR}$ 		&  $-11.78\pm0.4$ \kms&\\
P.A.               		& $100\arcdeg$ (fixed)&\\
$\log({\rm EM})$\tablenotemark{\scriptsize\dag}      	& $11.504\pm0.05$&\\
$M_\star$        		& $33$ \Msun (fixed) &\\
$i$     			& $144.8\arcdeg\pm0.1$\arcdeg &\\
$T_e$			&  $7000$  K (fixed)&\\    
$R_e$          		& $52\pm0.1$ au  &\\
$V_{\rm exp}$		& $10.9\pm3$ \kms&\\
$w_d$         			& $5.39\pm1$ au &\\
\enddata
\tablenotetext{$\scriptsize\dag$}{\small EM units: cm$^{-6}$ pc.}
\tablecomments{
\cchg{\small Using our notation, the inclination of the disk axis respect to the plane of the sky is $i-90^\circ$.}}
\end{deluxetable}

\begin{figure}
\includegraphics[width=\textwidth]{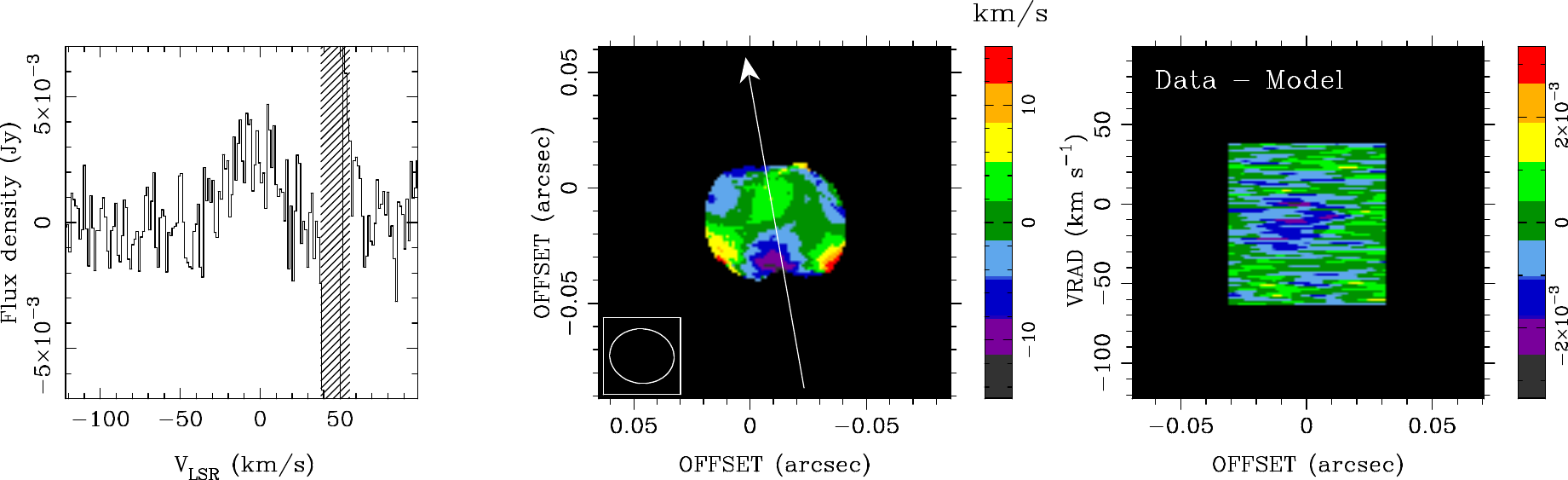}\\
\caption{Differences between the data and the best-fit model. Panels are arranged similarly to those in Figure  \ref{fig:disk} for comparison. 
\cchg{The shaded area shown in the first panel indicates the velocity range ignored by the fitting}. \label{fig:dif}}
\end{figure}

One of the most important parameters of the \hmyso\ system is the mass of the HMYSO. 
This mass is  degenerate with the inclination, resulting in higher mass estimations  for smaller angles.
There are several arguments that limit the possible mass or inclination of the \hmyso\ system. From the bolometric 
luminosity  of $1.5\times10^5$ $L_\sun$, 
we derive that the mass of this HMYSO is $M_\star=33$ \Msun\ using the stellar parameters from \citet{Davies2011MNRAS}.
The geometry of the outflow, jet, and outflow cavity also give some clues about the inclination of the disk, 
under the assumption that the outflow and the disk share the same axis. 
\citet{Guzman2016ApJ}, by comparing the jet lobes' proper motion velocity with the escape velocity of the star, derived that $i<154\arcdeg$ assuming $M_\star=15 M_\sun$. In general,  for $M_\star\le40 M_\sun$ and using this same argument, we obtain $i<160\arcdeg$.
\citet{Guzman2016ApJ} also show the NIR cavity observed toward the blue-shifted side of  \hmyso\, which has no extension toward the redshifted side of the jet. This  indicates that the line of sight toward the HMYSO  is not contained by the outflow cavity, and therefore the inclination angle is less than  $180\arcdeg-\theta_{\rm ap.}$, where $\theta_{\rm ap.}$ is the semi-aperture angle of the cavity.\footnote{From the NIR images we cannot derive a lower bound for $\theta_{\rm ap.}$. The projected aperture angle of the cavity is 45\arcdeg, that is, $2\theta_{\rm ap.}<45\arcdeg$.} Additionally,  the red- and blue-shifted halves of the CO bipolar molecular outflow associated with \hmyso\ are well separated \citep{Guzman2011ApJ}, indicating that the outflow is not near the plane of the sky or that $i$ is well above $90\arcdeg$.  Finally, we emphasize that due to the configuration of the outflow cavity  it is possible that the true bolometric luminosity  of \hmyso\ is  larger  than
that  given by \citet{Mottram2011AA}  by a factor of up to 3 due to the flashlight effect  \citep{Zhang2018ApJ}. This correction would increase  the mass estimate in about a 70\%.
 In Appendix A we present models assuming a HMYSO mass of 56 \Msun and 15 \Msun to encompass this uncertainty and the masses used also in previous literature \citep[e.g.,][]{Guzman2014ApJ}. 
\cchg{While in each table we provide the statistical error derived from the fitting, the dispersion of best-fit values  
can give a rough estimation of the systematic uncertainty derived using different assumptions. Specifically for the inclination, 
the range of HMYSO  masses  consistent with the luminosity constraints given above 
 is associated with a range of  inclinations which span at least $\pm20\arcdeg$ around the best-fit value given in Table \ref{tab:fit}.}


Figure \ref{fig:dif} shows the difference between the best-fit model and the data. Residuals of the moment 1 map 
are largest near the blue-shifted peak, while the pv-diagram shows some negative residuals (overestimation of the model) 
at the $\sim3\sigma$ level. 
Note also that whereas the projected elongation of a rotating 
disk is expected to occur   along the highest velocity gradient 
direction, the \hrl\ emission is more elongated in the perpendicular direction, i.e., along the jet direction. 
This is possibly due to expansion of the HC \hii\ region along  the polar axis (Section \ref{sec:wind}), which is not included in the model. 
This component may also produce the excess emission observed in the first panel of Figure \ref{fig:dif}. 

Another caveat of the expanding disk model is that while the moment 1 un-convolved model (middle-bottom panel in Figure \ref{fig:disk}) does indeed  produce the expected curvature of the locus of gas at the 
 $v_{\rm LSR}$ (the green hue velocities), after convolution the effect is  subtle (middle-middle panel). 
 \chg{A more distinct effect produced by the presence of the expanding radial wind is the inclination of the velocity 
 gradient respect to the longest projected axis of the disk. This effect is also subtle, but more noticeable even after convolution.
In fact, a disk without expansion but inclined with P.A.$=120\arcdeg$ gives only a slightly worse 
fit compared with that of  Table \ref{tab:fit}.}
Appendix \ref{sec:a-disk} shows the best-fit model results in this case.
Convolving the model with the clean beam is a simple approximation of the 
observational distortions introduced by the interferometer. A more rigorous approach
would be to perform simulated observations of the model matching the observational parameters. In Appendix B we show 
moment 1 maps of simulated observations of the model described above, the main finding being that the \cs-shape feature
is slightly better preserved compared with the simple convolution. \cchg{While  advantageous  in this regard, 
the computational costs of implementing  simulations to optimize the parameters put such  approach outside the scope of this paper.}

The residuals, caveats, and overall comparison between the model and the data 
indicate that the disk model  is certainly an oversimplification. However,  we think it captures important  features
of the \hrl\ and free-free continuum emission from \hmyso.  Considering that our current data is  able to
 resolve the velocity gradient with about three beams across, we refrain in this work, 
 from  including additional  sophistication to the model. The main difficulty we have now is the 
disagreement between the rotating compact system and the more extended emission.

{\subsection{Expansion of the HC \hii\   Region and Photo-Ionized Wind.}\label{sec:wind}}

One of the most obvious characteristics of the data, which is not well explained by a rotating disk, is the elongation of the source in the jet direction. This feature is also shown in \citet{Guzman2014ApJ}: their Source 10  (corresponding to \hmyso) seems elongated in the east-west direction, which corresponds roughly to the jet direction. This elongation is not readily  observed in the  0\farcs2 resolution data of \citet{Cesaroni2017AA}.

The elongation of the source can be explained by the expansion and consequently increase in size of the HC \hii\ region 
expected to occur in the directions where the material is less 
dense, that is,  towards the symmetry axis. The lower density is associated with 
a larger  distance that ionizing photons can travel before being absorbed.
 Density and opacity variations in the expanding gas  explain  the spectral index of the source and its variation in size with frequency \citep{Guzman2016ApJ}.

The dynamics of the HC \hii\ region, whatever its geometry, depends crucially 
on how its size compares  with the Bondi-Parker or sonic radius 
\begin{equation}
\rB = \frac{GM_\star }{2a^2}~~,\label{eq:rb}
\end{equation}
where $a = 9.08~\text{\kms}\sqrt{T_e/10^4~{\rm K}}$ is the sound speed of the ionized gas 
\citep{Sartorio2019MNRAS,Keto2007ApJ,Hollenbach1994ApJ}. Roughly speaking, hydrostatic solutions can be maintained 
by the gravity of the central star against the pressure gradient of the ionized gas if the size of the \hii\ region is
 less than $\rB$ (or $2\rB$,  \citealp{Hollenbach1994ApJ}).
If the ionized region extends beyond this size, then it can begin its classical pressure-driven expansion \citep[e.g.,][]{Dyson1997BOOK} 
which has been successfully used to describe the evolution of compact \hii\ regions \citep{Garay1999PASP}.

In the case of \hmyso\ and the best-fit parameters in Table \ref{tab:fit}, $\rB=$\chg{178 au}. 
We highlight that even in the lower limit of the stellar mass \chg{(15 \Msun\ suggested by \citealp{Guzman2014ApJ})}
\chg{we obtain $\rB=115$ au.}
Therefore, as shown in Figure \ref{fig:cont_m1_profile}, the ionized region we are probing likely lies inside the Bondi radius.
Because this size is too small to explain the  cm- and \mbox{3 mm}-wavelength data \citep{Guzman2016ApJ,Guzman2014ApJ}, 
there are two possible explanations: (i) the region extends in the polar directions up to the Bondi radius (and beyond) but the 
low brightness and large scale filtering precludes the detection of this emission, and (ii) the region we are observing is confined, 
but it does not absorb all of the ionizing radiation from the HMYSO. The remaining uv-radiation escapes through the polar regions to 
\chg{ionize} farther than 
$\rB$, extending the size of the ionized region. Observations covering intermediate wavelengths between 1.3 and 3 mm may help to determine what is the true geometry of the HC \hii\ region.


In the previous section, we argue that including a radial expansion component may help explain some characteristics of the \hrl\ moment 1 map. 
The best-fit parameters of this wind indicate it is  transonic, with Mach number $\mathcal{M}\gtrsim1$. The nature of this wind, however,  is not clear.
A first option is that it corresponds to  a pressure driven wind \citep{Parker1958ApJ},
 which would remain subsonic inside $\rB$.
However, at distances comparable with the size of the disk $R\approx 50$ au, 
the velocity derived for this wind is too small compared with the best-fit solution. Indeed, 
the Mach number at a distance $R$ from the origin is given by \citep{Lamers1999CUP}
\begin{equation}
\mathcal{M}\approx \left(\frac{R}{\rB}\right)^2\exp\left({-\frac{2R}{\rB}+\frac{3}{2}}\right)=0.20~~,
\end{equation}
which is \chg{lower} by a factor of at least $5$  compared with the derived $\mathcal{M}\gtrsim1$. 
We therefore rule out this wind coming from pressure driven expansion.

A second alternative is that this wind corresponds to photoionized gas being dragged by a stellar wind. 
These winds are common around high-mass stars but they are usually too tenuous to dominate the free-free and HRL emission.  
A stellar wind in \hmyso\ likewise helps to explain the relatively small best-fit disk width. 
Indeed, a strong stellar wind is expected to reduce the scale height of the photoionized disk, specially near a high-mass star
 \citep{Hollenbach1994ApJ}. 
However, the velocities of dragged photoionized winds are also predicted  to be somewhat higher than the ones 
fitted to \hmyso\ \citep{Yorke1996AA}.
A third alternative is that this wind is somewhat linked to the accretion process, specifically, that the wind corresponds to magneto-centrifugal acceleration. 
 This disk wind then would help the disk to lose angular momentum, facilitating the transport of material to smaller radii and eventually accrete into the HMYSO. This alternative has some support on the fact that \hmyso\ is associated with a fast, active  jet
and with evidence of accretion in the last hundreds of years. However,  the acceleration expected \citep[Equation (51)]{Tsinganos2007LNP} from such a mechanism also produces much larger 
velocities than the ones observed shortly  ($\le50$ yr) after  ejection.
At this point and with the current data, the true nature of the \hmyso\ wind remains speculative.

{\subsection{An Oppositely Rotating Molecular Envelope}\label{sec:rot}}

The first indication of rotation derived from angular momentum conservation and contraction in \hmyso\ was provided by \citet{Guzman2014ApJ},
who found that sulfur oxide molecule transitions traced a rotating molecular core of size $\sim4000$ au around the HMYSO. The rotation
axis was found to align well with the jet axis. 
No other molecular specie exhibit this kinematic feature \citep{Guzman2018ApJS} and a search for a more compact 
rotating structure in CH$_3$CN proved inconclusive at best \citep{Cesaroni2017AA}.

\begin{figure}
\includegraphics[width=1.05\textwidth]{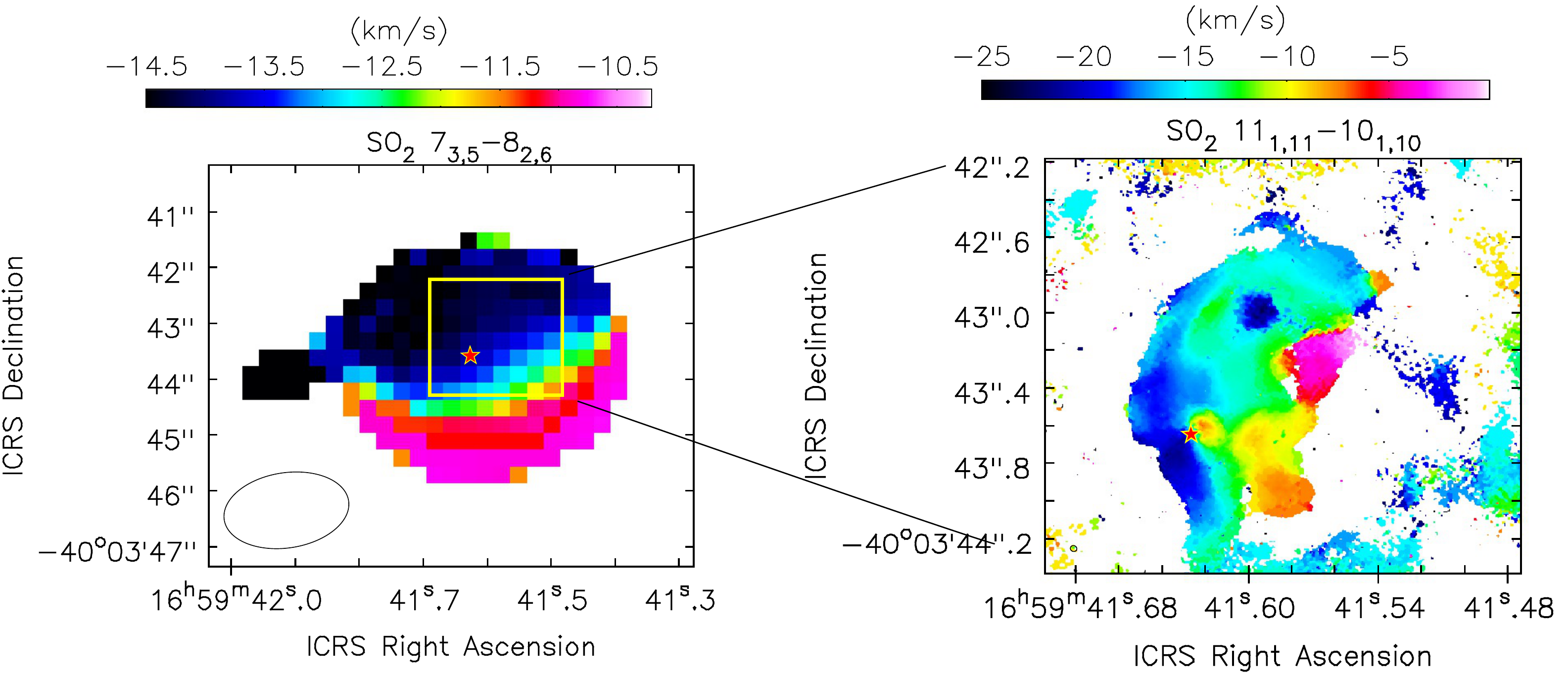}\\
\caption{Left panel: moment 1 of SO$_2$ $7_{3,5}\rightarrow8_{2,6}$ as presented in \citet{Guzman2014ApJ}. Square marks the zoomed-in region 
shown in the right panel. Right panel: moment 1 of the SO$_2$  $11_{1,11}\rightarrow10_{0,10}$ line. \textbf{Moment 1 is calculated using the line emission above $2\sigma$ and between $-36$ and $12$ \kms}. In both panels the star marks the position of 
\hmyso\ and the beam is shown in the lower-left corner. Note that the colorbars velocity range is different in both panels.\label{fig:m1so2}} 
\end{figure}

Figure \ref{fig:m1so2} shows the moment 1 of the SO$_2$ $11_{1,11}\rightarrow10_{0,10}$ transition ($E_u/k=60.4$ K) adjacent to the \hrl\ line (see \ Section \ref{sec:res})
compared with the 3 mm SO$_2$ $7_{3,5}\rightarrow8_{2,6}$ ($E_u/k=47.8$ K) moment 1 presented in \citet{Guzman2014ApJ}.
\chg{The SO$_2$ $11_{1,11}\rightarrow10_{0,10}$ moment 1 image was calculated using the CASA task \texttt{immoments} 
between $-36$ and $12$ \kms\ and considering only emission above the 2$\sigma$ level.}
While our data is heavily filtered out due to the lack of short baseline coverage, 
the moment 1  shows  a  velocity pattern roughly consistent with that traced by SO$_2$ at \mbox{3 mm}, 
although more aligned in the E-W direction and 
tracing a  higher velocity gradient. 
In fact, the SO$_2$ velocity pattern closer to the HMYSO presented in this work seems more aligned with the
jet and outflow. 

The most noticeable feature of the rotating pattern of SO and SO$_2$ at large scales is that it seems to run opposite to the rotation pattern of \hrl.  The radius of the SO$_2$ rotating core reported by
\citet{Guzman2014ApJ} is $\approx2100$ au, while  the mm HRLs probe sizes 40 times smaller.
Considering the difference in physical scales probed, it is not clear that the spin orientation
of the smaller scale structure should necessarily inherit that of the  larger structure. 
For example,  \citet{Bate2018MNRAS} finds in simulations
significant misalignment between the inner and outer parts of circumstellar disks, even in sources formed in relative isolation. 

Remarkably, this feature has already been observed also toward relatively ``evolved'' HMYSOs --- such as \hmyso\ --- in at least two other cases:
 in Orion Src I, \citet{Zapata2012ApJ} reported an elongated, $\sim2000$ au across structure in SiO rotating in the opposite direction compared to the  inner disk \citep{Hirota2017NatAs}; 
 and in \gds, \citet{Maud2018AA} found (also in SiO) that the lower velocity,  extended  structure  of
$\sim1000$ au across rotates in the opposite direction compared to the inner, $\sim240$ au disk \citep{Maud2019AA}.

\chg{Barring chance coincidence and combination of motions associated , there are  few theoretical physical explanations for these observed anti-alignments. 
First, in a crowded environment like a high-mass star forming region, it is possible that a combination of rotation and outflow motions associated with different sources could resemble an oppositely  rotating envelope \citep{Takakuwa2015ApJ}. Second, an oppositely rotating, expanding outer envelope is predicted by the effect of Hall currents in magnetized disks \citep{Tsukamoto2015ApJ}. These currents are expected to be important when the Hall scale becomes comparable with the size of the system \citep{Mininni2003ApJ}. For \hmyso\, assuming a completely ionized hydrogen plasma system of density $10^8$ cm$^{-3}$ and  velocities comparables with the Keplerian speed at a scale of 100 au (few tens of \kms), the Hall scale is extremely small (less than 1 m) even for a magnetic field as large as a milli-Gauss. It is possible, however, that the Hall current effect becomes more noticeable  farther from the HMYSO, where the ionization fraction becomes much lower.}
Finally, we note that there are similitudes between the chemistry in the three sources \chg{mentioned above}:  rotation is not probed 
by the typical hydrogenated hot-core tracer 
molecules, but rather   silicon and sulfur oxides, and water (the latter in Src I and \gds\ at least). 
\citet{Guzman2018ApJS} suggest that the chemical signature of the rotating core in \hmyso\ 
resembles somewhat that of envelopes of evolved stars, being  heavily 
influenced by the strong UV radiation. This view is also consistent with the results on the Orion Src I disk presented by \citet{Ginsburg2019ApJ}. 


{\subsection{Accretion onto \hmyso}\label{sec:acc}}

One important characteristics of \hmyso\ is that it is associated with a fast  protostellar jet \citep{Guzman2016ApJ}. This jet has excited at least 
two pairs  of roughly symmetrical ionized lobes whose proper motions  indicate
that one pair was probably ejected about  $450\pm140$ yr ago and the other  $\approx150\pm50$ yr ago.
The accretion rate associated is $\ge10^{-5}$ \Msun\ yr$^{-1}$.
Can the disk around \hmyso\ sustain these accretion rates?

First, the ionized mass in the disk is negligible, reaching only about $8\times10^{-6}$ \Msun, obtained by integrating the 
electron density $n_e$ (Equation \eqref{eq:ne}) times $\mu_e=1.3 m_{\rm H}$ 
(the mean mass per electron in a fully ionized \hii\ and  He\textsc{i} region). 
However, it is possible that most   of the disk mass
is neutral near the midplane, and that the ionized component consists of two layers on the surface of the disk. 
\chg{We can constrain the circumstellar mass of \hmyso\ by using the continuum excess over the free-free 
derived in Section \ref{sec:basic} (119 mJy) and the upper limit on the dust emission at 92 GHz of 11 mJy set by \citet{Guzman2014ApJ}.
For that, we assume that the thermal dust emission comes from an homogeneous  source covering a solid angle
$\Omega\approx 3\times10^{-3}$ arcsec$^2$ (Figure \ref{fig:contsat}), 
whose flux density is given by}
\begin{equation}
S_\nu=\Omega B_\nu(T_d)\left(1-\exp\left(-\frac{\kappa_\nu M}{\Omega d^2}\right)\right)~~,\label{eq:fdust}
\end{equation}
\chg{where $T_d$ is the dust temperature, $M$ its mass, and $d$ is the distance to the source. The dust opacity is given by 
$\kappa_\nu\propto \nu^\beta$  with $\beta=1.7$  and $\kappa_{\rm 231\,GHz}=0.00899$ cm$^2$ gr$^{-1}$
 is the dust opacity  \citep{Ossenkopf1994AA}  including a gas-to-dust mass ratio of 100. Taking into account 
 the flux limit at both frequencies and considering that $T_d$ should be less than the dust sublimation temperature \citep[typically 1500 K, e.g.,][]{Kama2009AA}
 we obtain $0.3\text{\Msun}\le M\le1.0 \text{\Msun}$ and $1500\,{\rm K}\ge T_d \ge 960\,{\rm K}$, respectively.}

%
\chg{A disk mass of 0.3--1.0} \Msun\ could  sustain accretion rates of $10^{-5}$ \Msun\ yr$^{-1}$ 
(the lower bound of the  accretion rate onto \hmyso, \citealp{Guzman2016ApJ}) 
for \chg{30,000--100,000 yr}, a time comparable with  estimations of the jet phase duration 
\citep[][approximately $5\times10^4$ yr]{Purser2016MNRAS,Guzman2012ApJ}. \chg{However, even if} the entire 
disk were accreted, it would only  affect minimally the mass of the HMYSO. 
Consistent with the disk not being too massive, we found no clear molecular counterpart of this disk in the rest of the frequency coverage of our data.


\chg{The estimation of the range of disk mass entails considering } two scenarios for \hmyso:
\begin{enumerate}[(i)]
\item{Accretion onto the HMYSO has essentially ended. The observed disk is a  remnant which 
 could sustain \chg{some  accretion}, as evidenced by the jet activity, which  cannot add a significant amount of mass to  the  HMYSO.}
\item{\chg{Accretion will resume in the future. We pose that accretion at this stage is very sporadic,} with bursts of accretion which  deplete a large fraction of the disk alternated between quiescent 
episodes with virtually no accretion, \chg{which would be the stage in which \hmyso\ is now}. The environment of the HMYSO  may replenish the disk and allow further accretion episodes \chg{in the future}.}
\end{enumerate}
Because there are  $\sim1000$ \Msun\ in the IRAS 16532$-$3959 molecular clump, 
which is in global contraction \citep{Guzman2011ApJ}, scenario (ii) is still plausible.
Accretion bursts associated with HMYSOs
have already been observed \citep[e.g.,][]{Caratti-o-Garatti2017NatPh,Hunter2017ApJ} and they can be substantial for the final mass of the star.

We propose the following picture for HMYSOs in a evolutionary stage similar to \hmyso. 
Just after an accretion burst, the HMYSO cannot adjust immediately, causing it to expand and cool down \citep{Hosokawa2009ApJ}. 
The lack of ionizing radiation allows material at $\lesssim10^4$ au  to approach  the HMYSO and replenish the disk more easily. 
However, if the timescale for disk replenishment 
becomes longer than the Kelvin-Helmholtz contraction timescale of the last accretion burst, the HMYSO will start 
to produce ionizing photons. This ionizing radiation is not effective  in stopping accretion in the immediate vicinity of the young star, but it can 
disperse the material on larger ($10^4$ au) scales \citep{Kuiper2018AA}, effectively halting further replenishment of the accretion disk. 
Therefore, that the cycle described in  (ii) perdures or becomes scenario (i) depends on 
the environment much farther away than the size of the disk, and on how the ionizing radiation will affect this environment. 

This very dynamic picture of accretion may  help to explain some of the observed variability of HC \hii\ regions \citep{Galvan-Madrid2008ApJ,DePree2014ApJ,DePree2015ApJ}.
A sufficiently massive  accretion burst may completely 
choke a young  HC \hii\ region, and the HMYSO will start again a new cycle of expansion and contraction. Likewise, 
the picture may help to understand  why HMYSOs with apparently little material in their
close  environment (suggesting more evolved sources)
 such as Orion Src I \citep{Plambeck2016ApJ} and IRAS 13481$-$6124 are not associated with \hii\ regions. In this view, 
 the immediate environment was possibly cleared out in the last  accretion burst, and the star is still contracting.  
 
This idea may provide a picture which explains  the oppositely rotating large scale cores described in Section \ref{sec:rot}.
In principle, if the disk is mostly consumed after an accretion burst, 
new replenishing material  should not necessarily form a disk with the same orientation or plane.
However, we must consider that the environment of the HMYSO is not 
homogeneous, but it has already been shaped by the previous outflows. In this way, the history of accretion shaping the environment
makes it likely that the new material  will accrete through a similar 
plane as their predecessor. The argument does not preclude  new material from having an opposite rotation.
 
 {\subsection{A likely binary system.}\label{sec-bin}}
High-mass stars commonly form in binary systems \citep[e.g.,][]{Chini2012MNRAS}. 
In the case of \hmyso, until now there has been  no  evidence that it is a binary. 
However, the detection of Source 10a suggests that this young star may be a stellar companion of 
the HMYSO. 
\chg{Alternative hypotheses are that Source 10a is far away from \hmyso, and only close in projection, or that  it  is a high-velocity fly-by young star.} 

Considering that Source 10a  is at a projected separation of 690 au from \hmyso, we correct this distance by $4/\pi$ assuming 
a random orientation between these sources, obtaining 880 au.
To determine whether a binary system with this separation is plausible in the center of the  IRAS 16562$-$3959 cluster environment, 
we compare the orbital energy of the  purported binary 
with the average energy of stellar encounters \citep[Chapter 8]{gady}. 
To evaluate the latter, we assume that the 3D velocity dispersion $\sigma$ of the young stars  
is the same as the observed velocity dispersion of the gas. 
Measurements of $\Delta v$ of the gas  toward IRAS 16562$-$3959  range between 4 and 7 \kms\ depending on the
molecular tracer \citep{Guzman2018ApJS,Miettinen2006AA,Bronfman1996AAS}, so, we 
use $\Delta v=5.5$ \kms (or $\sigmatD=\sqrt{3}\Delta v/2.35 = 4.05$ \kms) as an average value.
The criterion which defines a so-called hard binary in this environment is 
\begin{equation}
\frac{GM m}{r} > m_a\sigmatD^2~~,\label{eq-heggie}
\end{equation}
 where $r$ is the binay separation, $M$ and $m$ are the masses of both stars, and $m_a$ the typical clump stellar mass. 
 If the system does not fulfill \eqref{eq-heggie}, then it is prone to disruption by gravitational
  interaction with other cluster members. Since the distances involved are small compared with the  clump size, we 
   ignore the perturbing effect of the gas mass potential. 
 Further assuming that the mass of Source 10a is comparable to the typical  individual mass  of 
 the stellar component of IRAS 16532$-$3959, then Equation \eqref{eq-heggie} simply means that the orbital velocity around \hmyso\ should be 
 larger than the velocity dispersion of the clump. 
 
 \chg{Using $r=880$ au and $M=33$ \Msun, we obtain that the orbital velocity is 5.8 \kms,} larger than 
   $\sigmatD=4.05$ \kms. We conclude that \hmyso\ and Source 10a likely form a binary system. 
   The location  from the HMYSO makes Source 10a having formed from disk fragmentation
    rather unlikely. Only recently ALMA has 
provided the opportunity to witness how binaries containing high-mass stars form. In two \chg{recent} studies, 
\citet{Ilee2018ApJ}  finds that a high-mass star has a companion that  is likely a low-mass protostar (as in \hmyso), 
while \citet{Zhang2018NatAs} find a binary system in which both stars are massive and are associated with HC\hii\ regions.

{\section{Conclusions.}\label{sec:con}}
We present high spatial resolution  observations of \hmyso,   made using  ALMA,  
targeting the \hrl\ HRL and the 1.3 mm continuum emission. 
The main conclusions
of this study can be summarized as follows:
\begin{enumerate}
\item{The \hrl\ HRL  reveals a $\sim100$ au ionized disk around \hmyso, rotating perpendicularly to the previously detected ionized jet.
}
\item{The \chg{sense of} rotation observed in the ionized gas is opposite to that detected in sulfur oxides at larger scales.}
\item{The disk has sustained accretion and jet ejection episodes during the past hundreds of years. 
We estimate the disk mass  to be \chg{in the 0.3--1.0 \Msun\ range}, suggesting it is not relevant for the final mass of \hmyso. This disk may 
sustain an accretion rate 
of $10^{-5}$ \Msun\ yr$^{-1}$ \citep{Guzman2016ApJ} for  \chg{30,000--100,000} yr, comparable with estimates of the jet phase duration. 
 }
\item{We propose that the accretion onto \hmyso, and sources alike, is sporadic with strong bursts of accretion which consume a large fraction of the disk. 
We speculate that \hmyso\ is actually  between these accretion episodes.}
\item{The \hrl\ moment 1 map suggest the presence of a radial, expanding transonic wind in addition to rotation.}
\item{We report in this work   the detection of Source 10a, a likely low-mass YSO. This source is probably a binary stellar companion of \hmyso.}
\end{enumerate}

\acknowledgments
{Authors thank an anonymous referee for detailed comments which improved this work.
This paper makes use of the following ALMA data: ADS/JAO.ALMA\#2018.1.00385.S. ALMA is a partnership of ESO 
(representing its member states), NSF (USA) and NINS (Japan), together with NRC (Canada), MOST and ASIAA (Taiwan), and KASI (Republic of Korea), 
in cooperation with the Republic of Chile. The Joint ALMA Observatory is operated by ESO, AUI/NRAO and NAOJ. 
P.S. acknowledges partial support from a Grant-in-Aid for Scientific Research (KAKENHI Number 18H01259) of Japan Society for the Promotion of Science (JSPS).
L.A.Z. acknowledges financial support from CONACyT-280775, and UNAM-PAPIIT IN110618 grants, M\'exico. 
G.G.  acknowledges support from ANID Project AFB 170002. 
This work has made use of data from the European Space Agency (ESA) mission
{\it Gaia} (\url{https://www.cosmos.esa.int/gaia}), processed by the {\it Gaia}
Data Processing and Analysis Consortium (DPAC,
\url{https://www.cosmos.esa.int/web/gaia/dpac/consortium}). Funding for the DPAC
has been provided by national institutions, in particular the institutions
participating in the {\it Gaia} Multilateral Agreement. This paper made use of information from the Red MSX Source survey database at \url{http://rms.leeds.ac.uk/cgi-bin/public/RMS_DATABASE.cgi} which was constructed with support from the Science and Technology Facilities Council of the UK.
Portions of the analysis presented here made use of the Perl Data Language (PDL) developed by K. Glazebrook, J. Brinchmann, J. Cerney, C. DeForest, D. Hunt, T. Jenness, T. Lukka, R. Schwebel, and C. Soeller and can be obtained from \url{http://pdl.perl.org}.
}
\facilities{ALMA}
\software{CASA \citep[][]{McMullin2007ASPC}, Perl Data Language \url{http://pdl.perl.org}}
\clearpage
\appendix
\section{Disk Models with Different Parameters}\label{sec:a-disk}

\chg{We show best-fit results for the disk model using different set of parameters. 
Models A.1 shows  the best-fit results assuming that the  disk width  increases with radius as $w_d(r)=2H(r)$, with $H(r)$ as in the prescription of \citet[Eq.\ (2.2)][]{Hollenbach1994ApJ}.
Models A.2 and A.3 show the best-fit models assuming a mass of 15 and 56 \Msun, which span conservatively the uncertainty due to the flashlight effect and the range of 
masses given in the literature for \hmyso.
Models A.4 to A.6 show the fitted model with the same masses as in models A1 to A3,  but with P.A.$=120$\arcdeg.}

\begin{deluxetable}{lllllll}
\rotate
\tablecaption{Disk best-fit parameters \label{tab:a-fit}}
\tablecolumns{7}
\tablehead{
\colhead{Parameter} 	& \multicolumn{6}{c}{Best-fit values} \\
\colhead{~} 	& \colhead{A.1} & \colhead{A.2} & \colhead{A.3} & \colhead{A.4} & \colhead{A.5} & \colhead{A.6}
}
\startdata
$v_{\rm LSR}$ 							& $-11.91\pm0.2$ \kms		& $-11.71\pm0.4$ \kms		& $-11.78\pm0.2$ \kms		&  $-11.75\pm0.4$ \kms		& $-11.66\pm0.4$ \kms		& $-11.78\pm0.4$ \kms	\\ 
P.A.               							& $100\arcdeg$ (fixed)		& $100\arcdeg$ (fixed)  		& $100\arcdeg$ (fixed)		& $120\arcdeg$ (fixed)		& $120\arcdeg$ (fixed)  		& $120\arcdeg$ (fixed)	\\
$\log({\rm EM})$\tablenotemark{\scriptsize\dag}& $11.504\pm0.02$			& $11.475\pm0.03$			& $11.583\pm0.02$			& $11.510\pm0.04$			& $11.460\pm0.03$			& $11.574\pm0.06$		\\
$M_\star$        							& $33$ \Msun\  (fixed)		& $15$ \Msun\  (fixed)		& $56$ \Msun\  (fixed)		& $33$ \Msun\  (fixed)		& $15$ \Msun\  (fixed)		& $56$ \Msun\  (fixed)	\\
$i$     								& $144.8\arcdeg\pm3\arcdeg$	& $140.2\arcdeg\pm4\arcdeg$	& $153.4\arcdeg\pm7\arcdeg$	& $144.5\arcdeg\pm1\arcdeg$&$137.3\arcdeg\pm3\arcdeg$ & $151.8\arcdeg\pm1\arcdeg$	\\
$T_e$								&  $7000$  K (fixed)			& $7000$  K (fixed)			& $7000$  K (fixed)			&  $7000$  K (fixed)			& $7000$  K (fixed)			& $7000$  K (fixed)		\\    
$R_e$          							& $51.9\pm0.1$ au  			& $52.3\pm0.1$ au			& $53.3\pm2$ au			& $52.0\pm0.2$ au  			& $53.9\pm0.1$ au			& $53.4\pm0.1$ au		\\
$V_{\rm exp}$							& $16.5\pm3$ \kms			& $12.2\pm3$ \kms			& $13.3\pm2$ \kms			& $0.6\pm3$ \kms			& $1.8\pm3$ \kms			& $-0.4\pm3$ \kms		\\
$w_d$          							& \nodata		 			& $3.62\pm0.6$ au			& $6.83\pm0.8$ au			& $4.69\pm0.7$ au			& $2.96\pm0.4$ au			& $6.3\pm1$ au		\\
\enddata
\tablenotetext{$\scriptsize\dag$}{EM units: cm$^{-6}$ pc.}
\tablecomments{\cchg{\small Using our notation, the inclination of the disk axis respect to the plane of the sky is $i-90^\circ$.}}
\end{deluxetable}

\begin{figure}
\includegraphics[width=\textwidth]{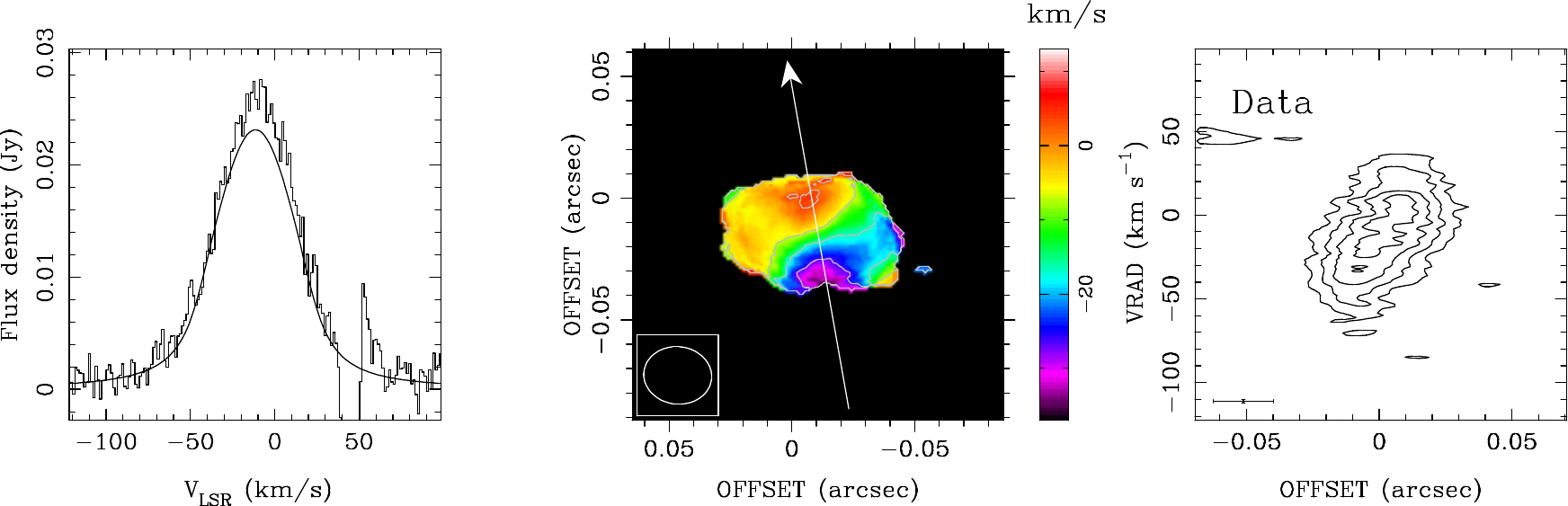}\\
\includegraphics[width=\textwidth]{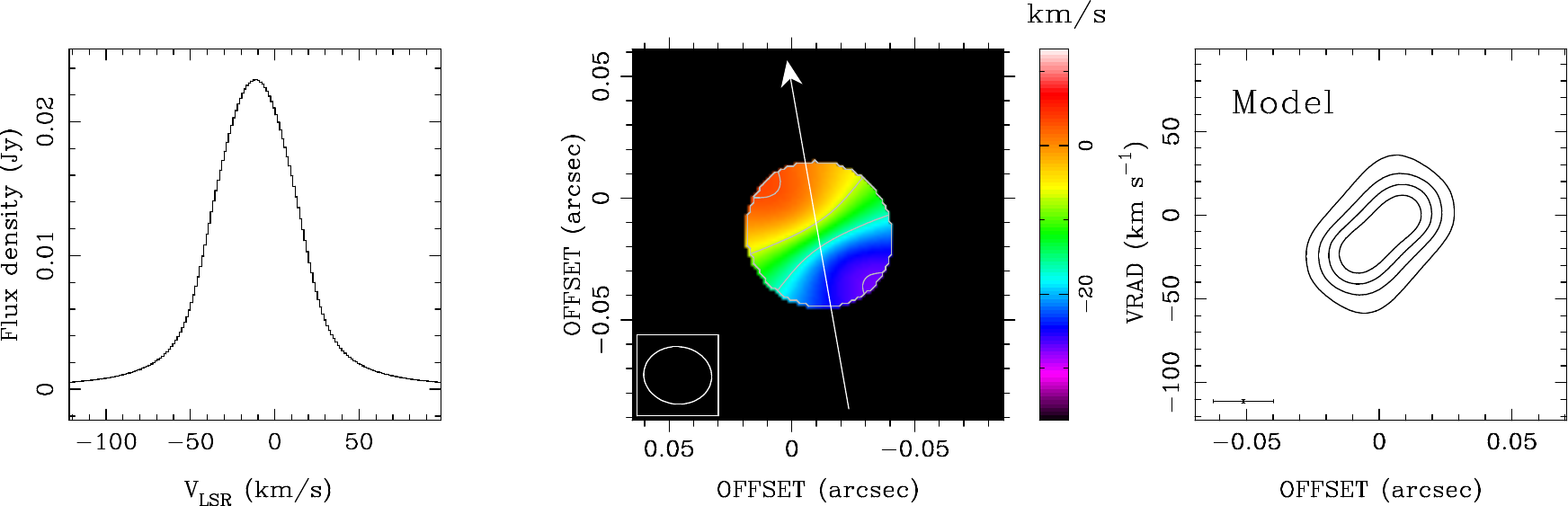}\\
\includegraphics[width=\textwidth]{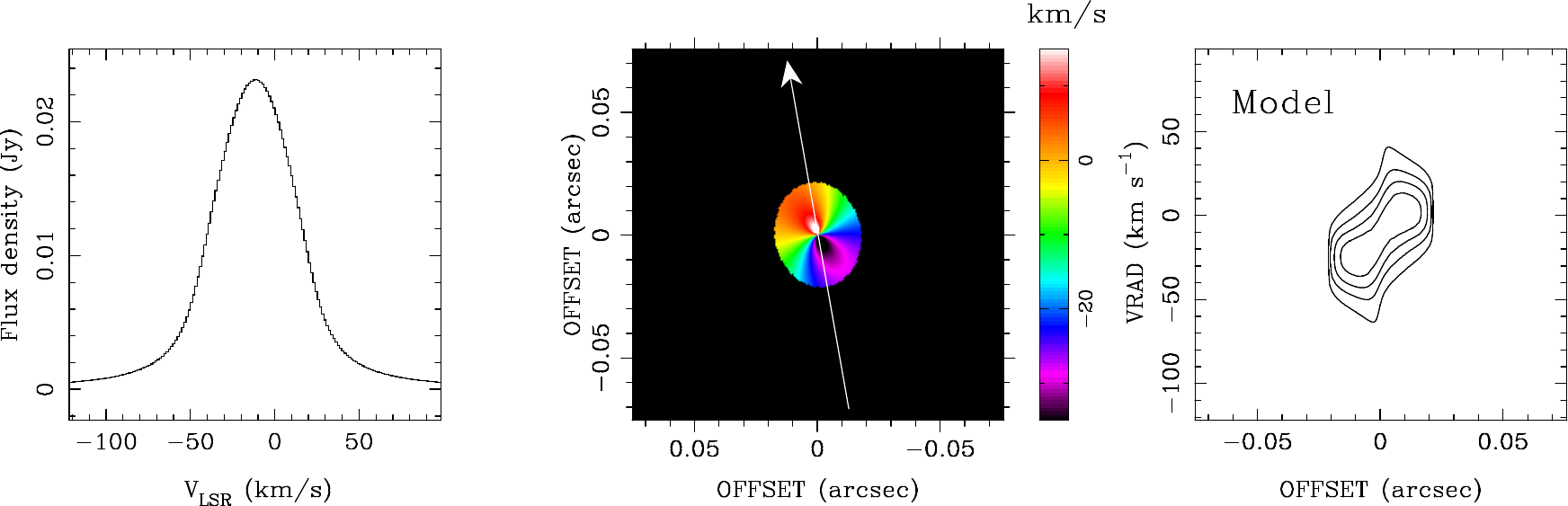}%
\caption{Model A.1. First row, from left to right: panels show the integrated spectrum, moment 1 (five contour levels between $-35$ and $15$ \kms),  
and pv-diagram (contour levels at 20, 40, 60, and 80\% of the  peak $=6.9$ mJy beam$^{-1}$), respectively. The arrow in the second panel shows 
the size and orientation of the pv-diagram cut.
Second row:  best-fit 
model (Table \ref{tab:a-fit}) convolved with the beam. Model spectrum shown in the first  panel is the same as in the first panel of the first row. Contour levels 
in second and third plots are the same as the respective panels in the first row. 
Third row: best-fit model  not convolved with the beam.  \label{fig:a1}}
\end{figure}

\begin{figure}
\includegraphics[width=\textwidth]{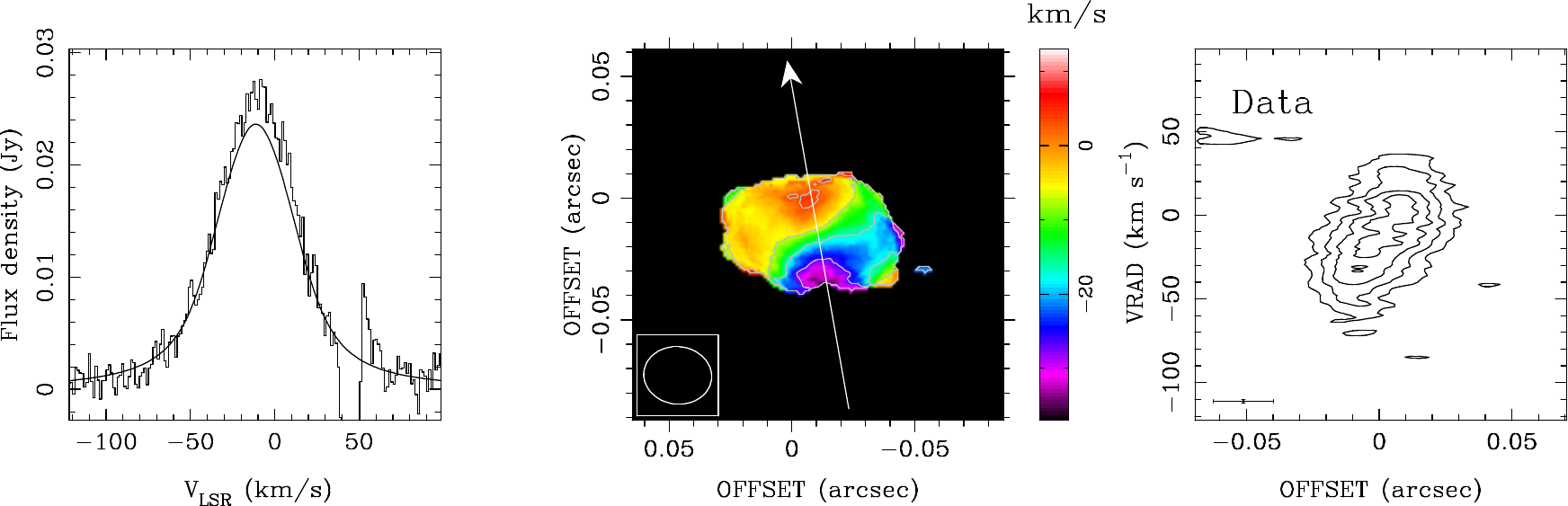}\\
\includegraphics[width=\textwidth]{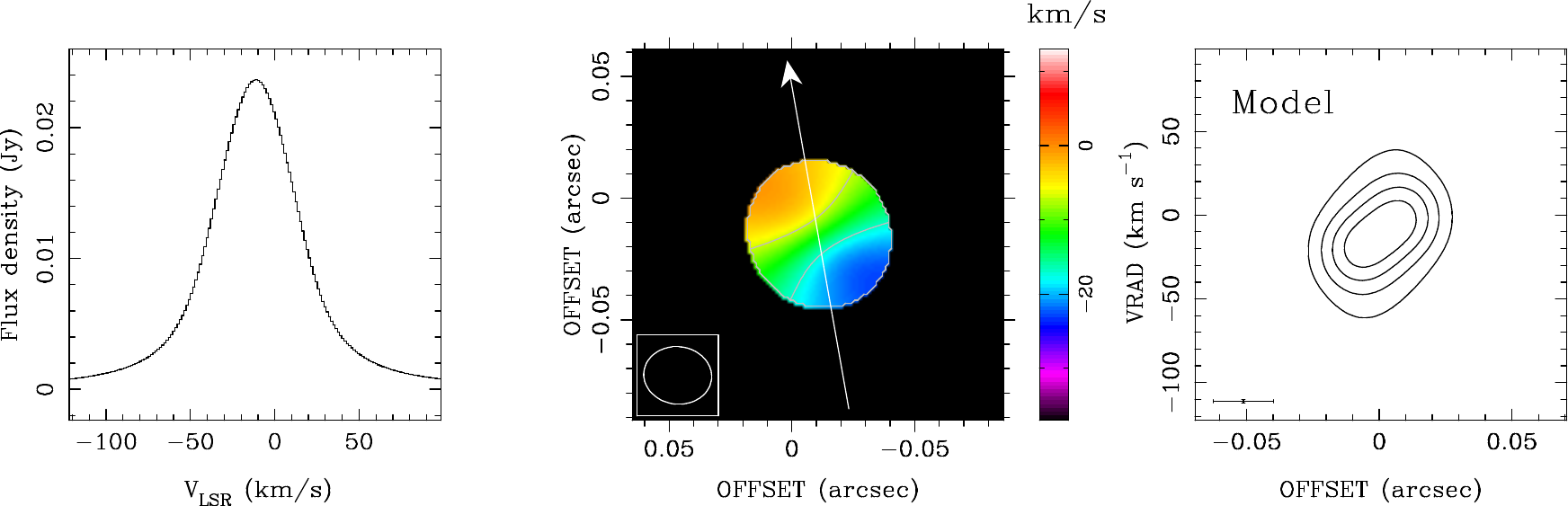}\\
\includegraphics[width=\textwidth]{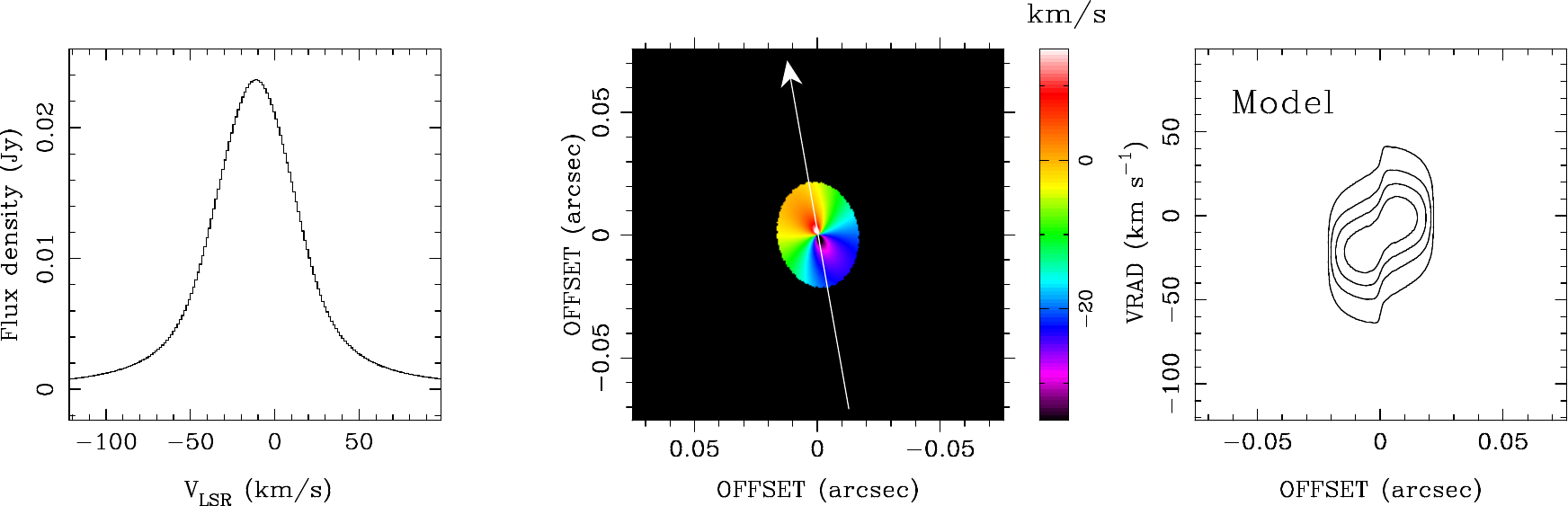}%
\caption{Model A.2. Panels, contours, and symbols same as in Figure \ref{fig:a1}. Parameters from Table \ref{tab:a-fit}.  \label{fig:a2}}
\end{figure}

\begin{figure}
\includegraphics[width=\textwidth]{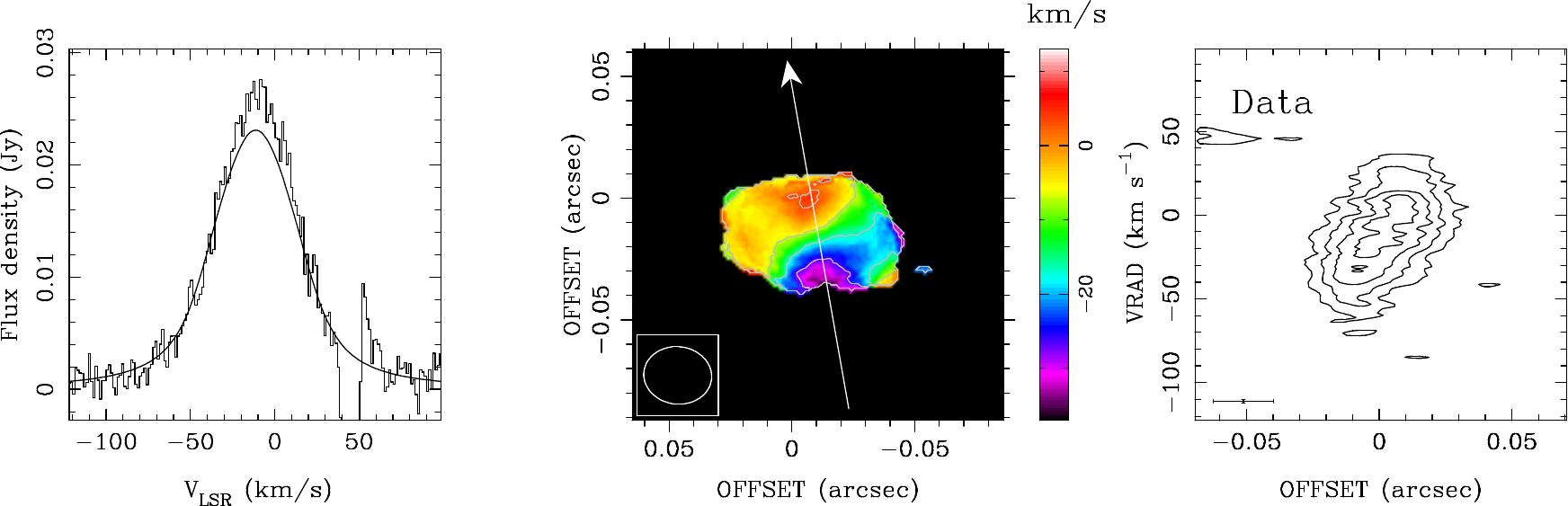}\\
\includegraphics[width=\textwidth]{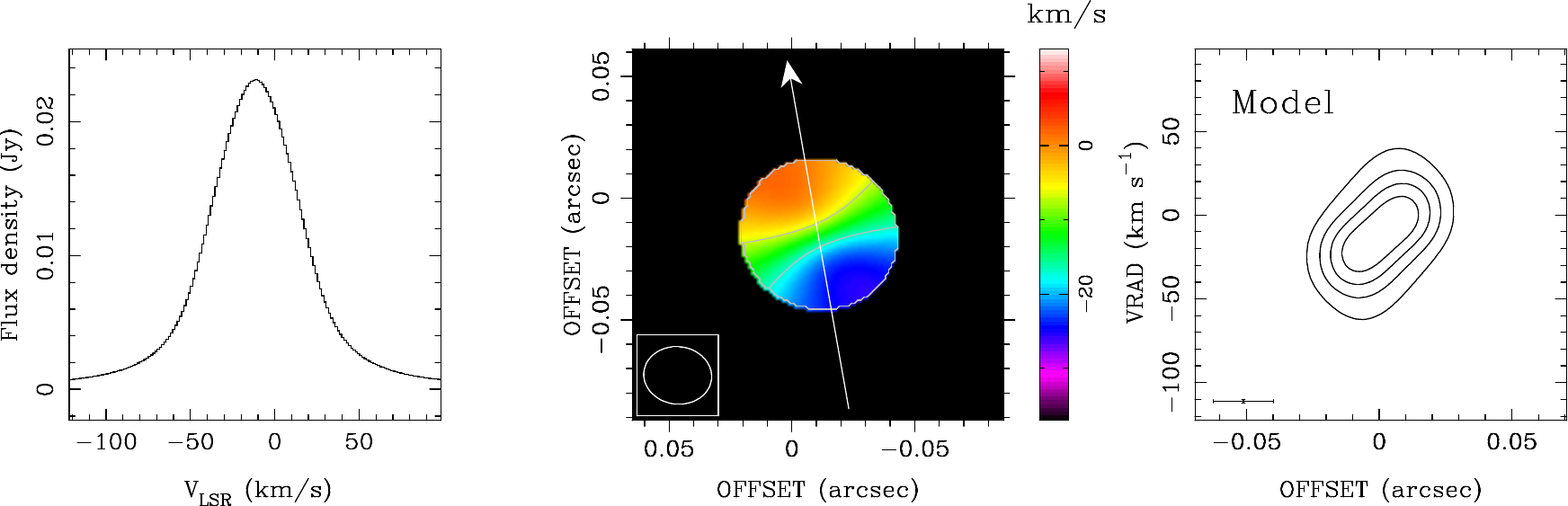}\\
\includegraphics[width=\textwidth]{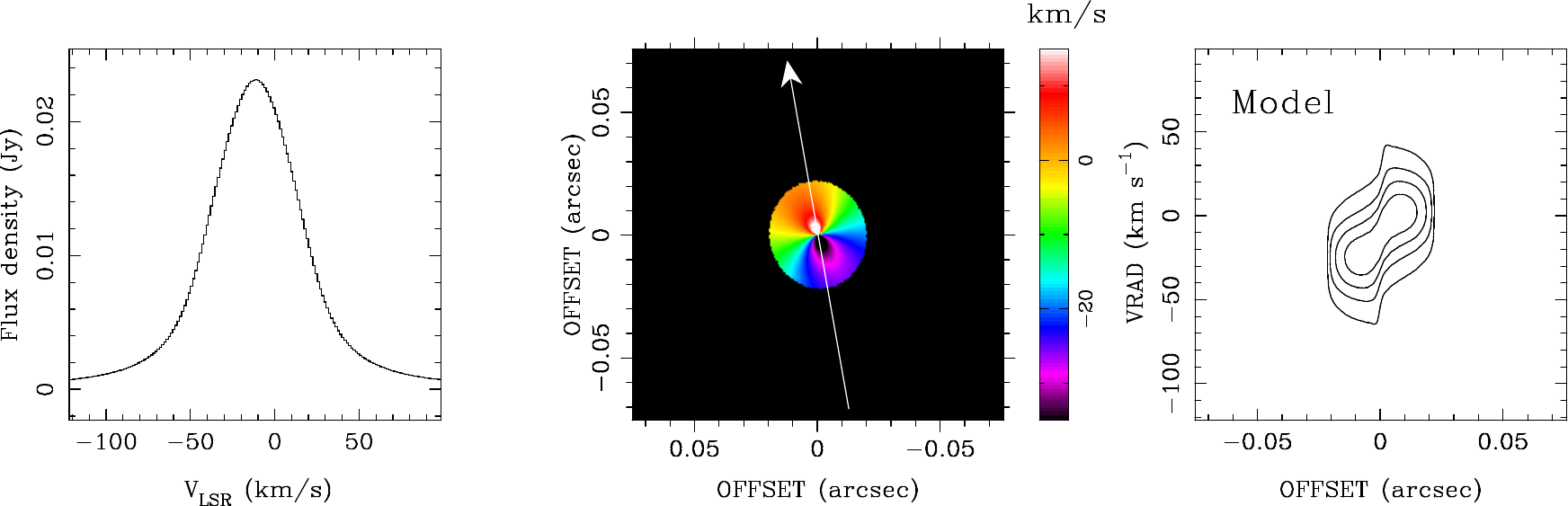}%
\caption{Model A.3. Panels, contours, and symbols same as in Figure \ref{fig:a1}. Parameters from Table \ref{tab:a-fit}.  \label{fig:a3}}
\end{figure}

\begin{figure}
\includegraphics[width=\textwidth]{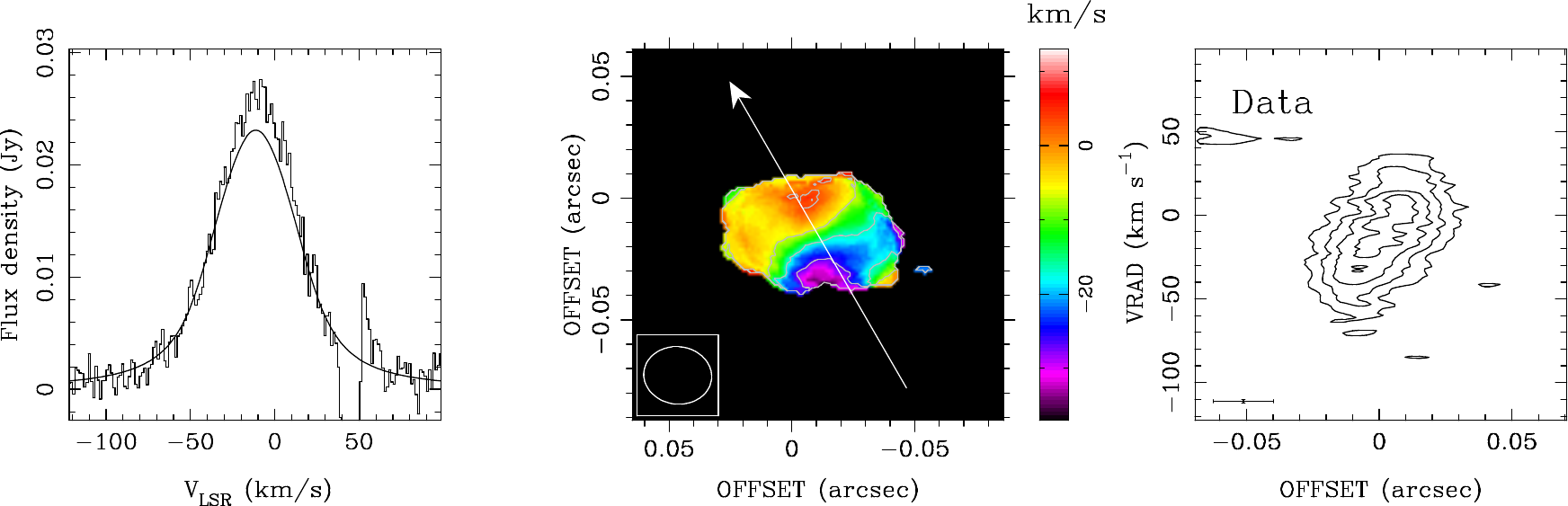}\\
\includegraphics[width=\textwidth]{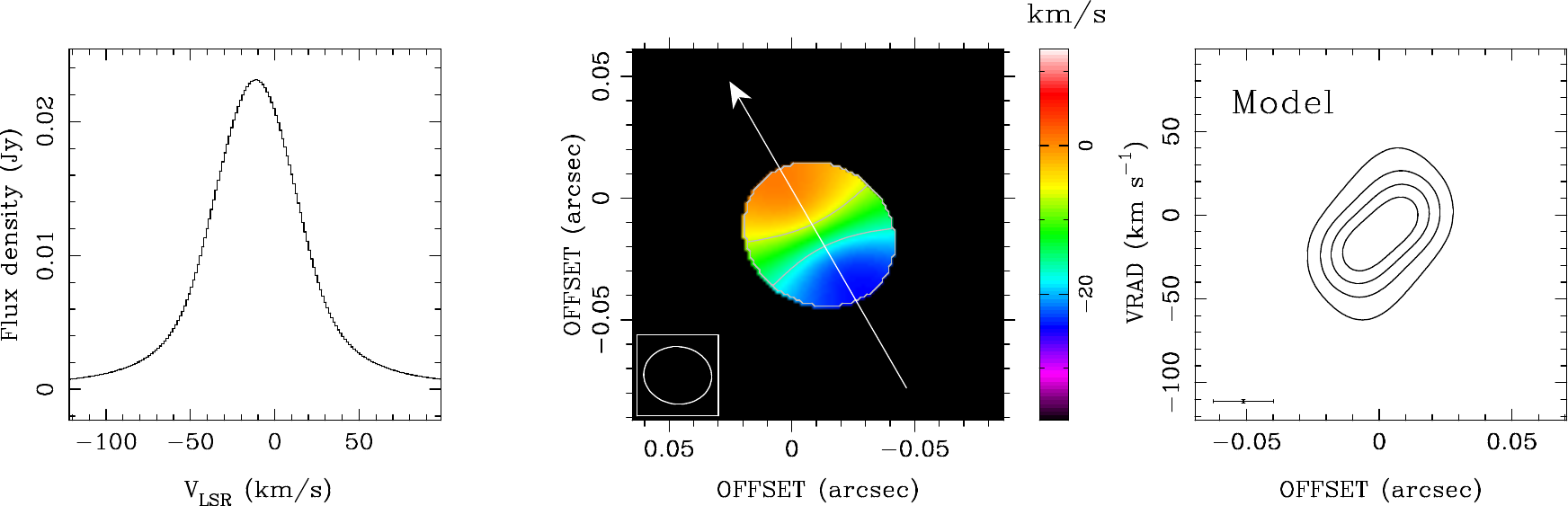}\\
\includegraphics[width=\textwidth]{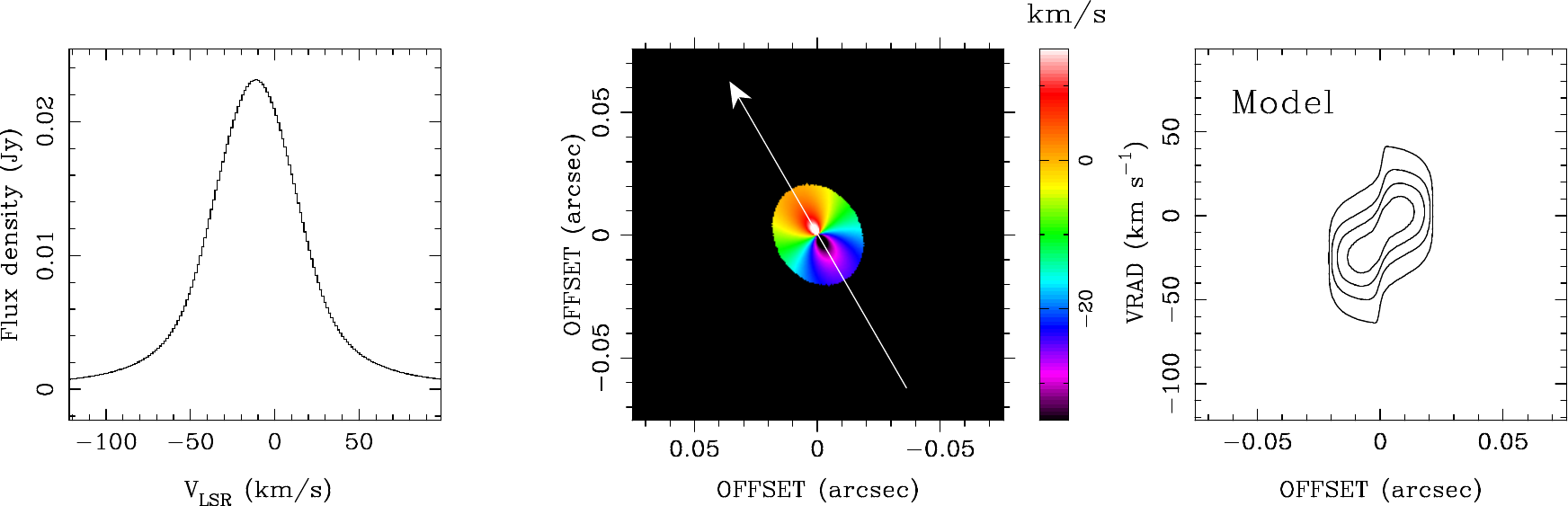}%
\caption{Model A.4. Panels, contours, and symbols same as in Figure \ref{fig:a1}. Parameters from Table \ref{tab:a-fit}.  \label{fig:a4}}
\end{figure}

\begin{figure}
\includegraphics[width=\textwidth]{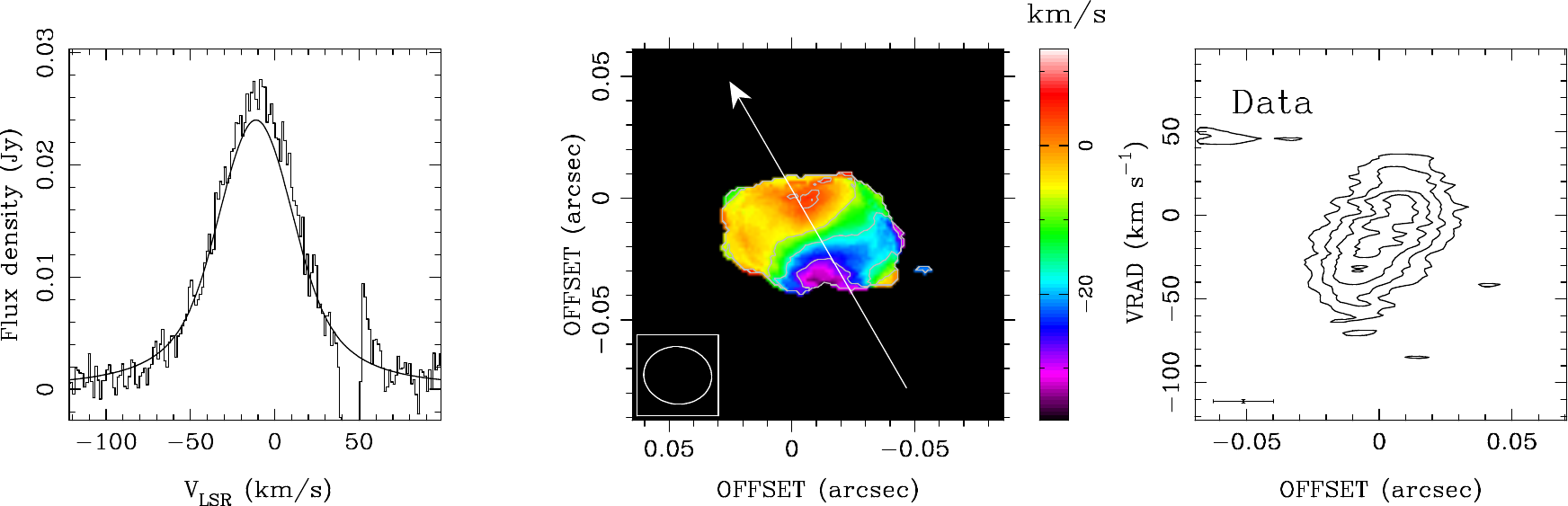}\\
\includegraphics[width=\textwidth]{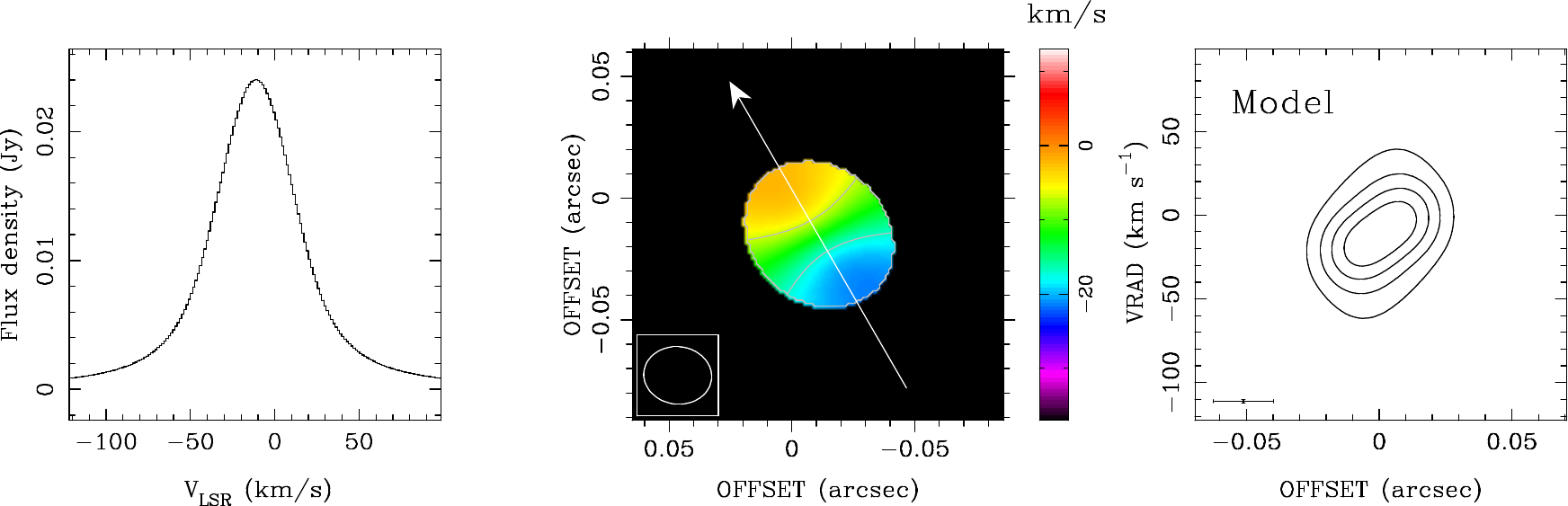}\\
\includegraphics[width=\textwidth]{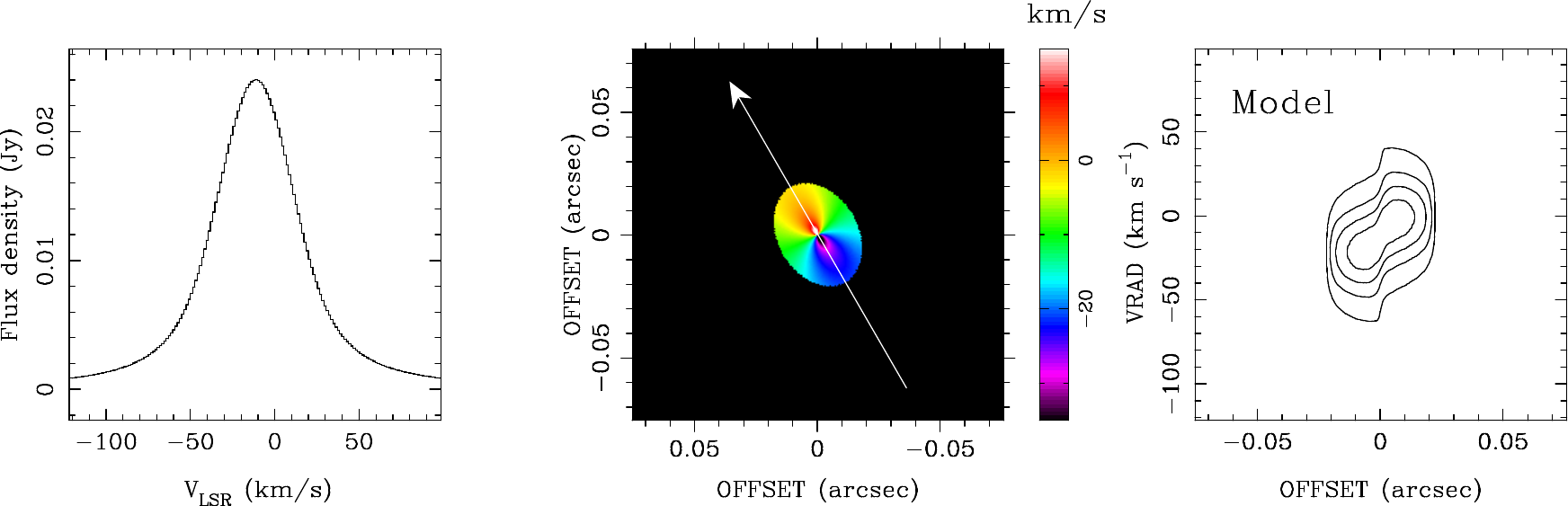}%
\caption{Model A.5. Panels, contours, and symbols same as in Figure \ref{fig:a1}. Parameters from Table \ref{tab:a-fit}.  \label{fig:a5}}
\end{figure}

\begin{figure}
\includegraphics[width=\textwidth]{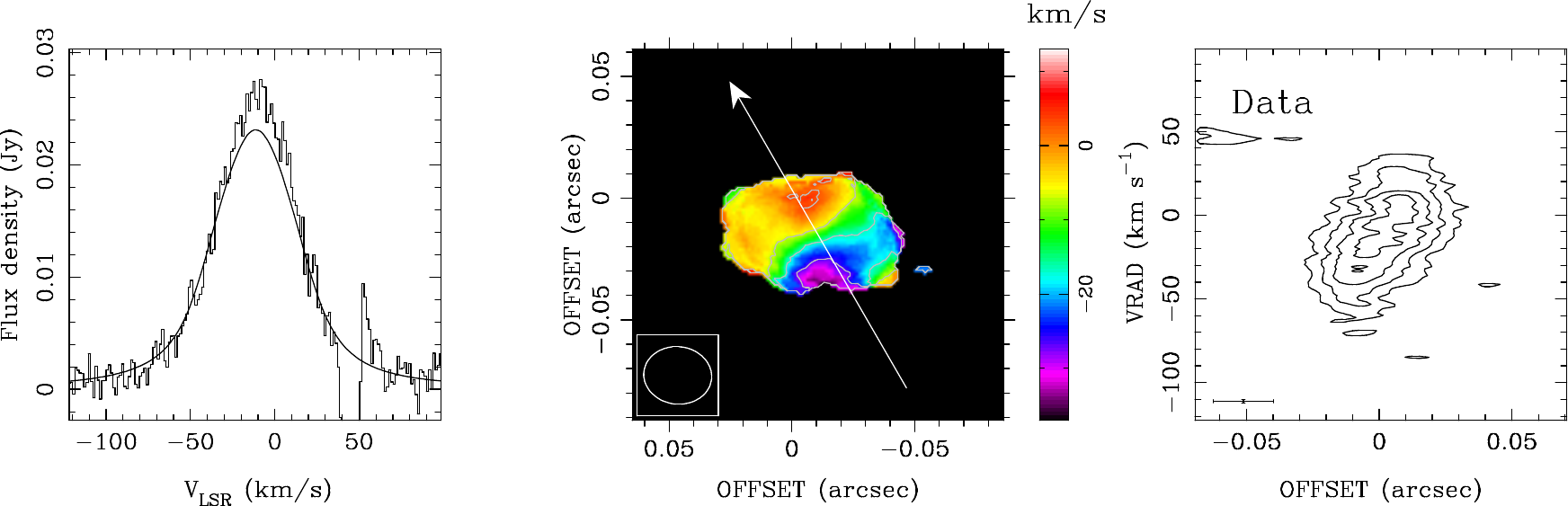}\\
\includegraphics[width=\textwidth]{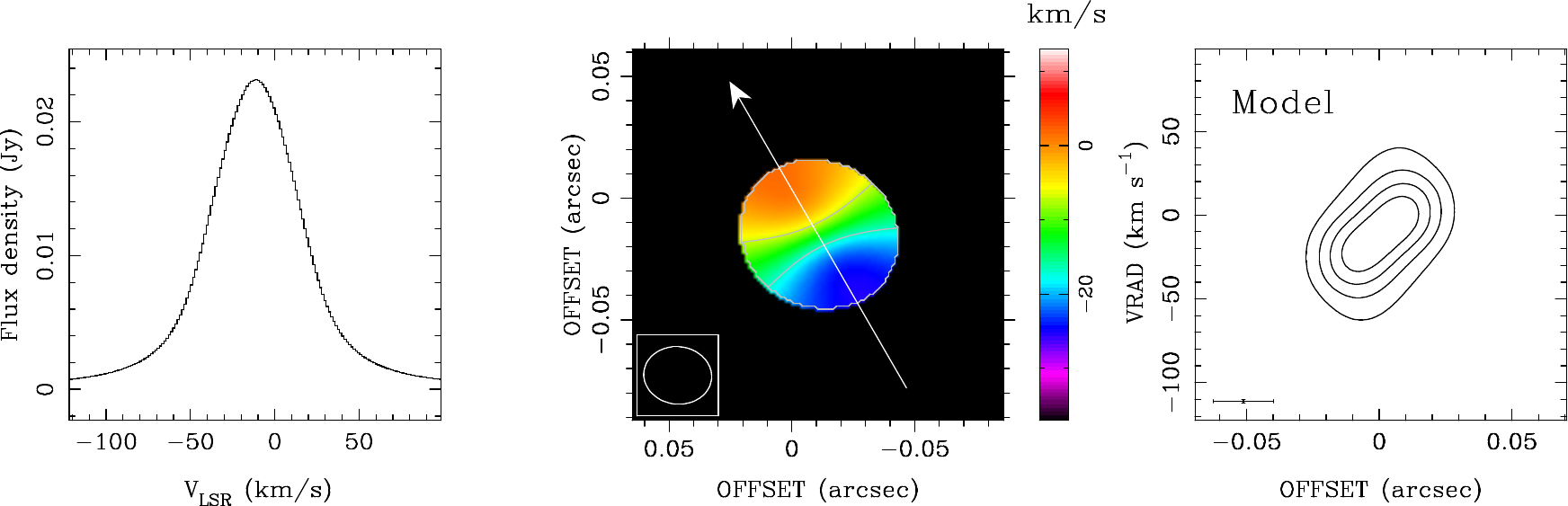}\\
\includegraphics[width=\textwidth]{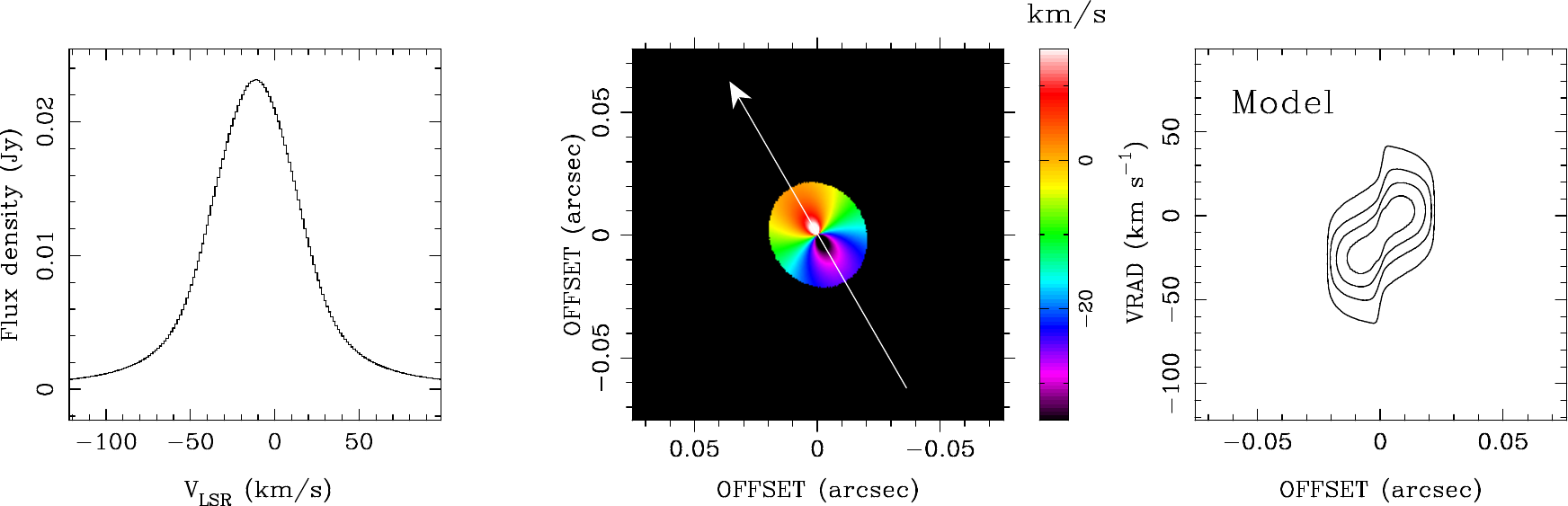}%
\caption{Model A.6. Panels, contours, and symbols same as in Figure \ref{fig:a1}. Parameters from Table \ref{tab:a-fit}.  \label{fig:a6}}
\end{figure}

\begin{figure}
\includegraphics[width=\textwidth]{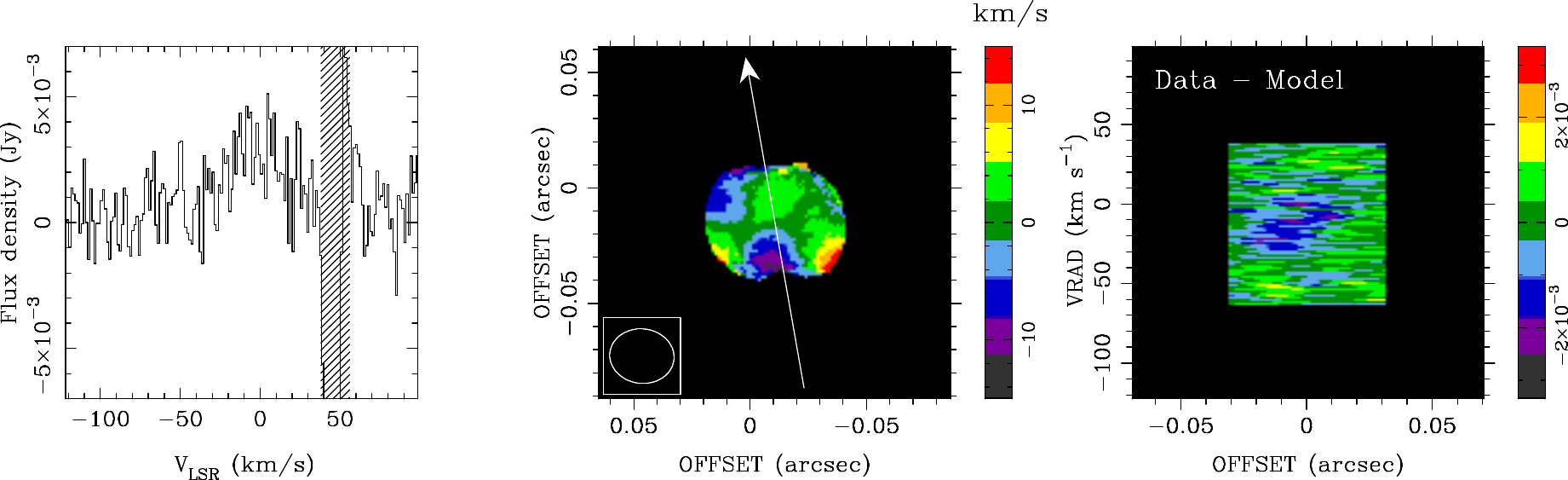}\\
\includegraphics[width=\textwidth]{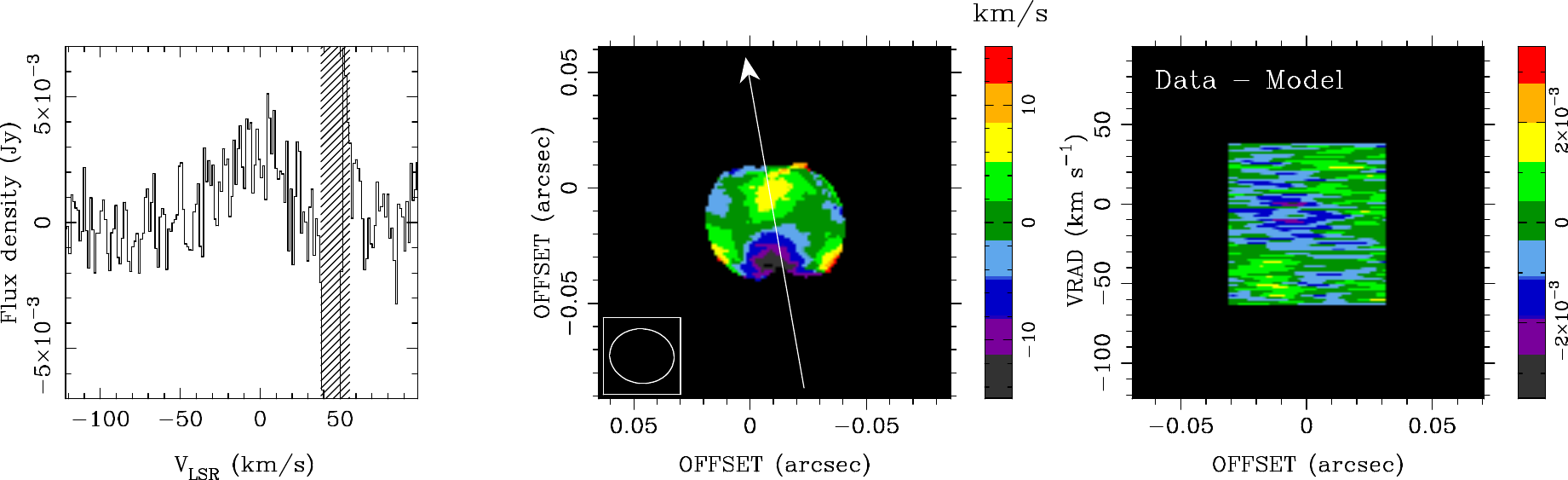}\\
\includegraphics[width=\textwidth]{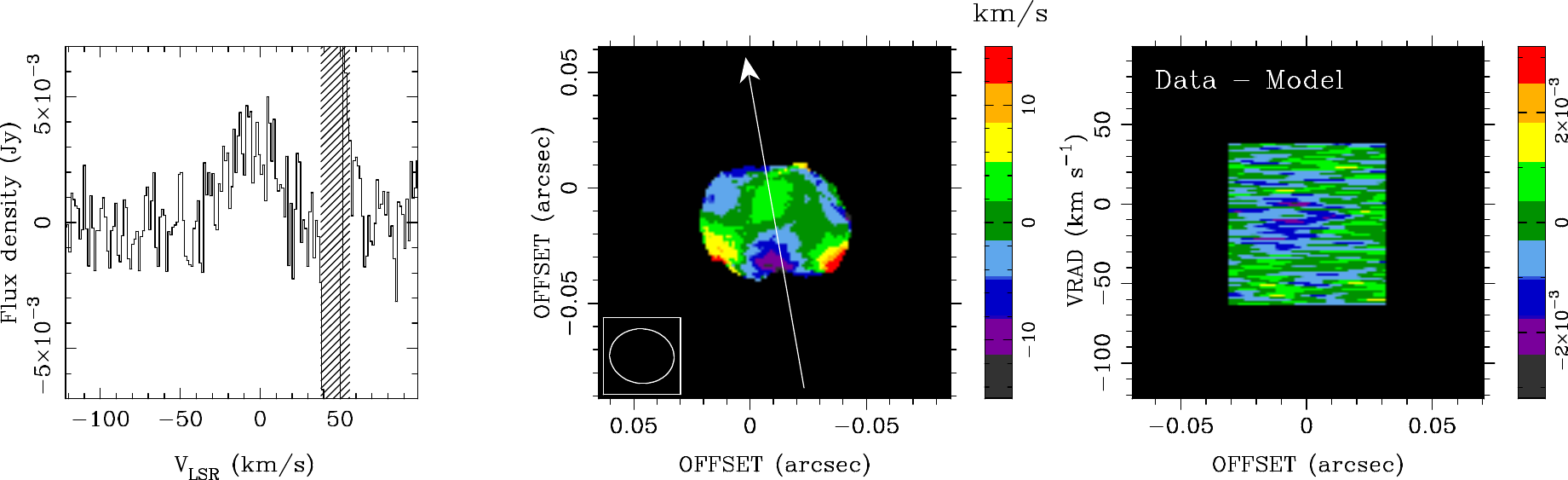}\\
\caption{From top to bottom, the three rows show the differences between the data and the best-fit models A.1, A.2, and A.3, respectively. 
Panels are arranged similarly compared with Figure \ref{fig:dif}.\label{fig:a123-dif}}
\end{figure}

\begin{figure}
\includegraphics[width=\textwidth]{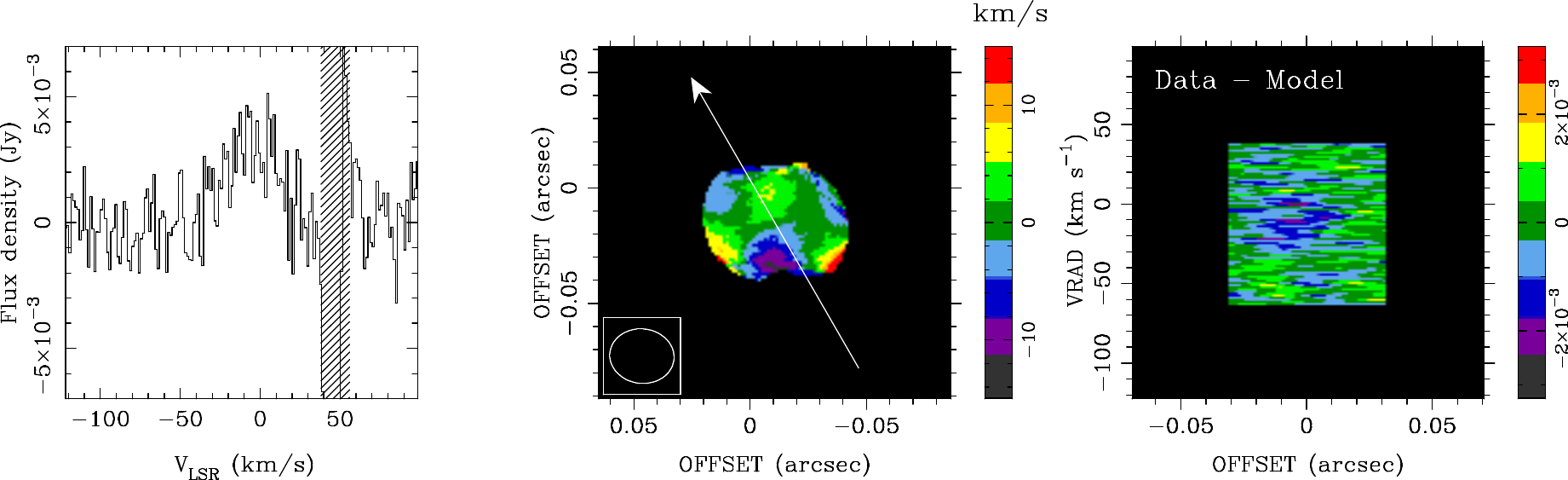}\\
\includegraphics[width=\textwidth]{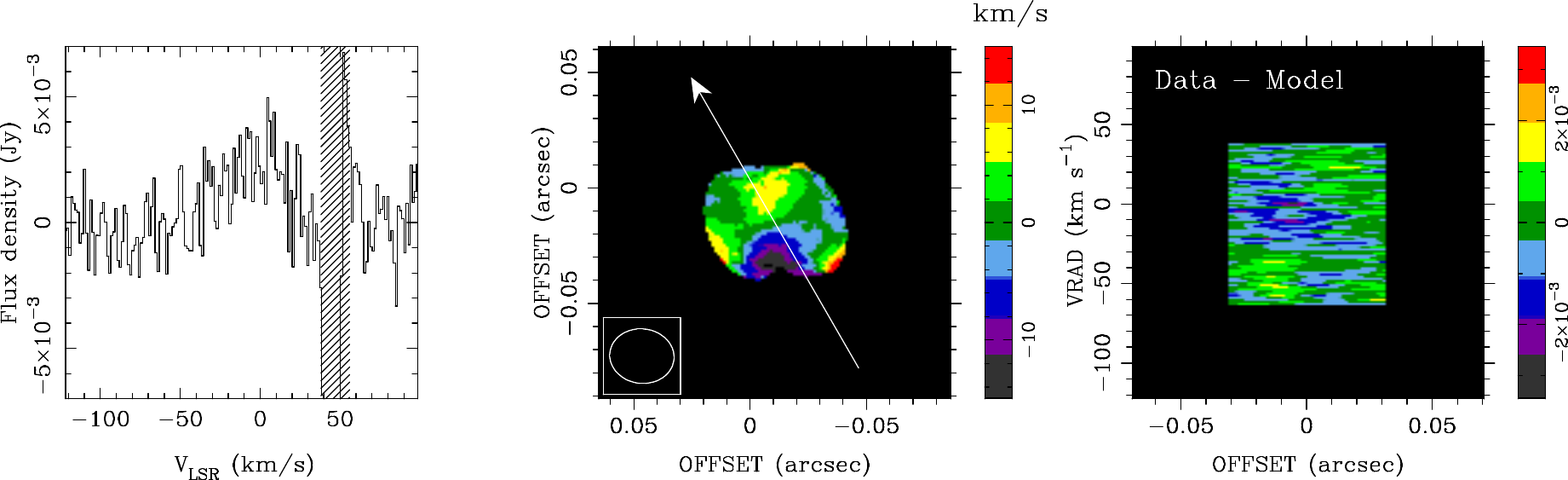}\\
\includegraphics[width=\textwidth]{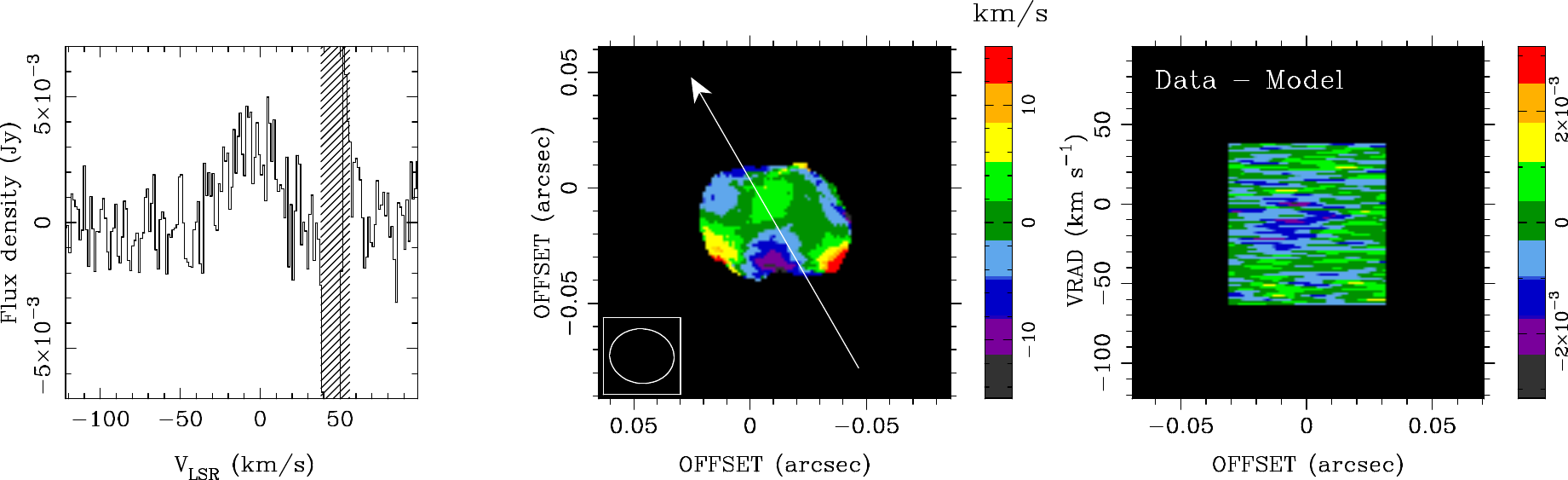}\\
\caption{ Same as Figure \ref{fig:a123-dif} but for models A.4, A.5, and A.6 in each row, respectively.\label{fig:a456-dif}}
\end{figure}

\clearpage
{\section{Simulated observations}}
\chg{In order to compare the data with the model and determine the best fit parameters, the 
approach followed in this paper (Section 4) is simply to convolve the model with the clean beam to introduce the effect of limited resolution. 
Whereas this process is relatively simple and computationally cheap, the clean beam is only an approximation of one aspect of the instrumental effect and observational uncertainties introduced by the telescope. In order to reproduce better the instrumental response, we   simulated observations of a theoretical brightness model (third row of Figure \ref{fig:disk}) using the CASA task \texttt{simobserve}. We created  the configuration antenna file for the two  execution blocks of our observations using the \texttt{buildConfigurationFile} task of the CASA extensions\footnote{\url{https://safe.nrao.edu/wiki/bin/view/ALMA/BuildConfigurationFile}}. After that, using the same LST range as with our visibilities, we run \texttt{simobserve} to create the simulated measurement sets.}

\chg{We restored the cubes to the image plane and deconvolved the simulated datacubes using the task \texttt{tclean} with the same parameters 
as described in Section 2. Given the width of the HRL, this process requires generating cubes of around 200 channels wide at least. Because of the 
very long baselines involved in our observations, adequately sampling the beam response and avoiding  aliasing entails  creating  large cubes of $11,000\times11,000$ pixels per channel. Partly because of this computational cost, we decided not to implement this approach in the optimization steps.}

\begin{figure}
\includegraphics[width=\textwidth]{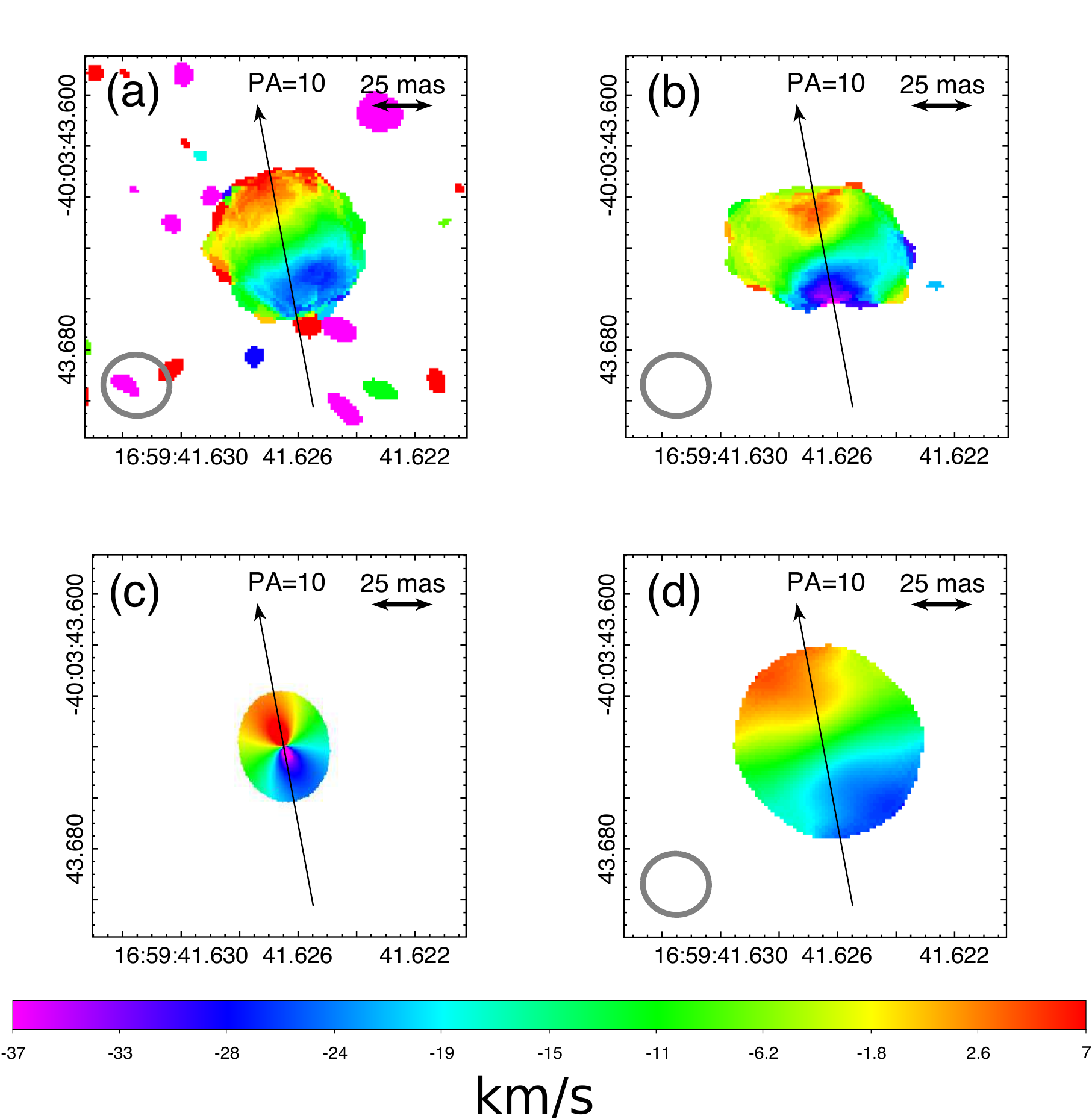}\caption{Moment 1 maps of the data and models. Panel (a): simulated deconvolved data with simulated thermal noise. Panel (b): \hrl\ data. Panel (c): theoretical model using parameters in Table \ref{tab:fit}. Panel (d): noiseless simulated deconvolved data.  Clean beam is shown in the lower left corner of panels (a), (b), and (d).\label{fig:sim}}
\end{figure}

\chg{Figure \ref{fig:sim} shows the result of the simulated observations of the best-fit model whose parameters are in Table \ref{tab:fit}. Panels (a) to (d) show the the moment 1 of the \texttt{simobserve} cube including simulated observational noise into the uv-data, the observed moment 1 map, the moment 1 of the model, and the moment 1 of the \texttt{simobserve} cube deconvolved from simulated noiseless uv-data, respectively. Panels (a) and (d) show that the \cs\ signature in the moment 1 seems to be better preserved in this simulation compared to the clean beam convolution. However, we note that this depends somewhat on the noise level up to which we can effectively calculate the moment 1; the \cs-shape feature being more conspicuous the less bright emission in the moment 1 calculation is included.}

\clearpage

\bibliography{bibliografia}
\bibliographystyle{aasjournal}
\end{document}